\documentclass{article}

\usepackage{arxiv}

\usepackage[utf8]{inputenc}                   
\usepackage[T1]{fontenc}                      
\usepackage{hyperref}                         
\usepackage{url}                              
\usepackage{booktabs}                         
\usepackage{amsfonts}                         
\usepackage{nicefrac}                         
\usepackage{microtype}                        
\usepackage[capitalise,noabbrev]{cleveref}    
\usepackage{graphicx}
\usepackage{subcaption}
\usepackage[authoryear]{natbib}
\usepackage{doi}
\usepackage{minted}
\usepackage{amstext}
\usepackage{multirow}
\usepackage[minted,breakable]{tcolorbox}
\usepackage{tikz}
\usetikzlibrary{decorations.pathreplacing}

\makeatletter
\def\maxwidth{\ifdim\Gin@nat@width>\linewidth\linewidth\else\Gin@nat@width\fi}
\def\maxheight{\ifdim\Gin@nat@height>\textheight\textheight\else\Gin@nat@height\fi}
\makeatother
\setkeys{Gin}{width=\maxwidth, height=\maxheight, keepaspectratio}

\newcommand{\approach}{MultiMend}

\title{\approach{}: Multilingual Program Repair with Context Augmentation and Multi-Hunk Patch Generation}

\date{}

\author{ \href{https://orcid.org/0000-0001-6596-3658}{\includegraphics[scale=0.06]{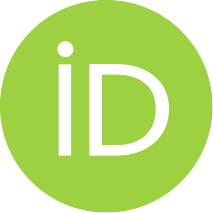}\hspace{1mm}Reza Gharibi}, Mohammad~Hadi Sadreddini, \href{https://orcid.org/0000-0002-9517-0541}{\includegraphics[scale=0.06]{figures/orcid.pdf}\hspace{1mm}Seyed~Mostafa Fakhrahmad}\thanks{Corresponding author}\\
  Department of Computer Science and Engineering and IT \\
	School of Electrical and Computer Engineering \\
	Shiraz University, Shiraz, Iran \\
	\texttt{gharibi@cse.shirazu.ac.ir, sadredin@shirazu.ac.ir, fakhrahmad@shirazu.ac.ir} \\
}

\renewcommand{\headeright}{}
\renewcommand{\undertitle}{}
\renewcommand{\shorttitle}{}

\hypersetup{
pdftitle={\approach{}: Multilingual Program Repair with Context Augmentation and Multi-Hunk Patch Generation},
pdfsubject={cs.SE, cs.LG},
pdfauthor={Reza Gharibi, Mohammad~Hadi Sadreddini, Seyed~Mostafa Fakhrahmad},
pdfkeywords={Automated program repair, Deep learning, Transformer, Retrieval-augmented generation, Code language model},
}

\begin{document}
\maketitle
\setcounter{footnote}{0}

\begin{abstract}
  Debugging software remains a labor-intensive and time-consuming process despite advances in testing and verification. Learning-based automated program repair (APR) has shown promise in reducing the effort of manually fixing bugs. However, existing techniques face several challenges, including language-dependent strategies, limited bug context utilization, and difficulties in handling bugs that span multiple locations in the code. This paper presents \approach{}, a multilingual learning-based APR approach designed to improve repair performance through language-independent context augmentation and multi-hunk patch generation. \approach{} fine-tunes a pre-trained code language model to generate bug-fixing patches. It embeds source code lines and applies retrieval-augmented generation to augment the usual function-based buggy context with relevant lines during patch generation. The approach also systematically constructs patches for multi-hunk bugs to extend the capabilities of single-hunk models and reduce the needed patch validations. We evaluate \approach{} on six benchmarks with 5,501 bugs covering four programming languages and compare it with state-of-the-art methods. Results show that \approach{} achieves competitive effectiveness and efficiency, fixing 2,227 bugs, of which 1,545 are identical to the developer's patch, and 121 are for multi-hunk bugs. Both context augmentation and multi-hunk patch generation contribute positively to these results. Overall, \approach{}'s contributions are promising and offer practical and effective techniques to enhance APR performance for real-world software maintenance.
\end{abstract}


\section{Introduction}
\label{sec:introduction}

Software bugs are inevitable consequences of the increasing complexity and rapid evolution of software systems. Despite advances in testing and verification techniques, debugging remains a labor-intensive process that consumes a substantial portion of software development and maintenance efforts. This not only slows down the development process but also increases the likelihood of introducing new bugs during the debugging process itself \citep{hamillAnalyzingPredictingEffort2017}.
Automated program repair (APR) has emerged as a promising approach to alleviate this burden, aiming to automatically fix bugs in software code. By doing so, APR techniques can potentially improve software reliability and accelerate development workflows \citep{legouesAutomatedProgramRepair2019}.
Recent APR approaches leverage learning-based methods, particularly sequence-to-sequence models trained on large datasets of buggy-fixed pairs, to propose patches for unseen bugs. This data-driven paradigm enables models to learn repair patterns by capturing complex relationships between the buggy code and its fix \citep{zhangSurveyLearningbasedAutomated2023}.

Despite these advancements, existing APR techniques still face several challenges. Learning-based APR tools typically rely on the buggy lines of code and their context to generate patches. The quality and scope of this context are critical for the model's ability to extract repair ingredients or capture repair examples necessary for generating accurate patches. In many cases, this context is often limited to the immediate lines around the buggy hunk (a set of consecutive code change lines) or the function enclosing it \citep{gharibiT5APREmpoweringAutomated2024, jiangCURECodeAwareNeural2021, lutellierCoCoNuTCombiningContextaware2020}. However, some bugs need more than their surrounding context, and the relevant information might be far from the buggy locations. Enriching the model's input by augmenting the buggy context helps the repair model reason about code dependencies and relationships, thereby improving the quality of generated patches. Some tools use an external database of bug-fixing examples to retrieve context-augmenting examples \citep{nashidRetrievalBasedPromptSelection2023, wangRAPGenRetrievalAugmentedPatch2023}. When external examples are not used, many approaches are tailored to specific programming languages and use language-dependent strategies to incorporate project-specific information into the context \citep{chenSequenceRSequencetoSequenceLearning2019}.

Another limitation of learning-based APR techniques lies in the focus on single-hunk patches, which address defects that need a fix in a single contiguous location. However, many real-world bugs often span multiple locations within the same function or file, or even across multiple files in the code, requiring multi-hunk patch generation. While some learning-based tools target specific classes of multi-hunk defects, there usually is not an explicit systematic approach to how they generate a full patch that resolves all related hunks of a bug. Only a few approaches propose explicit methodologies for constructing a unified patch for all the locations of multi-hunk bugs \citep{gharibiT5APREmpoweringAutomated2024, huangEmpiricalStudyFineTuning2023, liDEARNovelDeep2022, yeITERIterativeNeural2024}.

In this paper, we introduce \approach{}, a novel approach for multilingual program repair that leverages context augmentation and multi-hunk patch generation to effectively repair complex bugs. \approach{} employs a language-independent context augmentation strategy based on retrieval-augmented generation (RAG) \citep{gaoRetrievalAugmentedGenerationLarge2024, lewisRetrievalAugmentedGenerationKnowledgeIntensive2020} to enrich the input with relevant examples and repair ingredients. Unlike existing approaches, \approach{} retrieves context directly from the same buggy file, eliminating the need for external bug-fixing databases. Specifically, for each buggy hunk, we embed lines outside the immediate surrounding context using a vector embedding model. Then, during inference, we use a retriever to retrieve relevant lines based on their similarity to the buggy hunk. These retrieved lines provide additional ingredients or repair examples for constructing patches, ensuring the utilization of relevant contextual information for the bug \citep{barrPlasticSurgeryHypothesis2014}.

We also propose a systematic approach to generate patches for multi-hunk bugs, as the combination of patches for multiple hunks increases the size of the search space and complicates the repair process. Building on ideas from T5APR \citep{gharibiT5APREmpoweringAutomated2024} and ITER \citep{yeITERIterativeNeural2024}, \approach{} first attempts to identify patches that can be applied uniformly across all hunks, targeting scenarios where identical changes resolve multiple related hunks simultaneously. Additionally, \approach{} validates patches one hunk at a time while retaining partial fixes that address subsets of failing tests to eventually combine them to form a full patch for the multi-hunk bug. This strategy allows efficient convergence on plausible multi-hunk patches with fewer candidate validations. It also enables models designed to fix single-hunk bugs to address multi-hunk bugs as well.

We build \approach{} following T5APR's framework to leverage its efficient multilingual capabilities and evaluate it on six widely used benchmarks, including Defects4J \citep{justDefects4JDatabaseExisting2014}, QuixBugs \citep{linQuixBugsMultilingualProgram2017}, Codeflaws \citep{tanCodeflawsProgrammingCompetition2017}, BugAID \citep{hanamDiscoveringBugPatterns2016}, BugsInPy \citep{widyasariBugsInPyDatabaseExisting2020}, and RunBugRun \citep{prennerRunBugRunExecutableDataset2023}. The results demonstrate that \approach{} is effective at generating patches for real-world bugs and is competitive with existing state-of-the-art APR tools. Across the evaluated benchmarks, \approach{} generates 3,121 plausible patches, where 2,227 are correct when considering only the first plausible patch, and 1,545 are identical to the ground-truth developer patch. It also correctly fixes 121 multi-hunk bugs.

Our main contributions are summarized as follows:
\begin{itemize}
  \item Introduction of \approach{}, a multilingual APR tool that leverages context augmentation with the ability to generate patches for multi-hunk bugs (\cref{sec:approach}).
  \item A language-independent context augmentation method using RAG to enrich the repair input with relevant information from the buggy files during inference (\cref{subsec:augmentation-generation}).
  \item A systematic strategy for generating cohesive multi-hunk patches, extending the applicability of the APR tool to repair bugs requiring fixes in multiple non-contiguous locations (\cref{subsec:patch-validation}).
  \item Extensive evaluations of \approach{} on multiple benchmarks across four programming languages, demonstrating its generalizability and effectiveness (\cref{sec:experimental-setup,sec:results}).
\end{itemize}

We discuss related work in \cref{sec:related-work} and conclude the paper in \cref{sec:conclusion} with suggestions for future directions.

\section{Approach}
\label{sec:approach}

\begin{figure}
  \centering
  \includegraphics{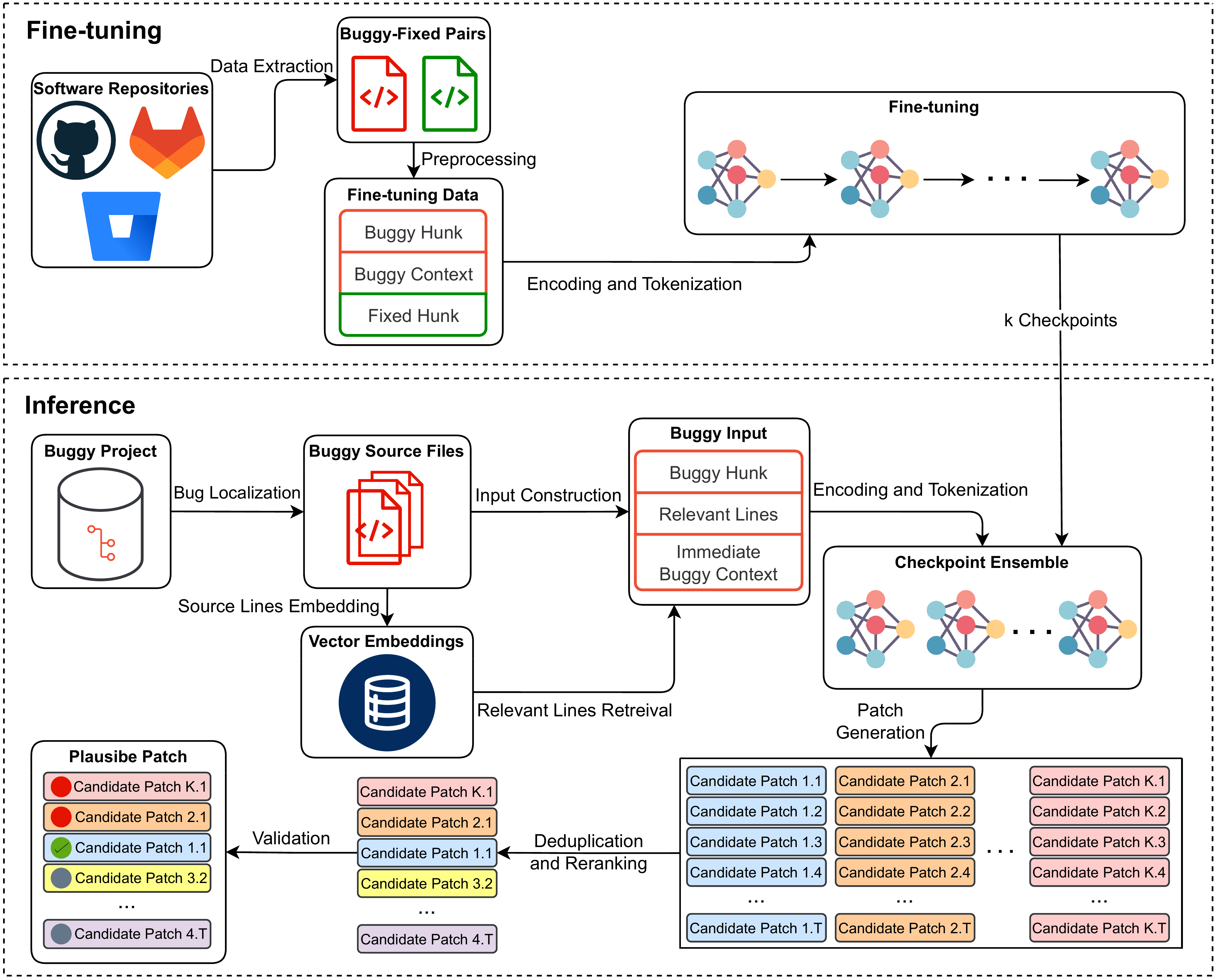}
  \caption{Overview of \approach{}.}
  \label{fig:arch}
\end{figure}

\approach{} consists of two major stages: Fine-tuning and inference. For fine-tuning, data is extracted from historical buggy-fixed pairs of various software repositories. Then, the pre-trained CodeT5 \citep{wangCodeT5IdentifierawareUnified2021} model is fine-tuned using the multilingual dataset of buggy hunks, their contexts, and fixed counterparts. Throughout the fine-tuning process, we select several checkpoints or snapshots of the model at different steps for the inference stage.

In the inference stage, the buggy project is analyzed, and the location of the bug is identified to construct an input for patch generation. The input consists of the buggy hunk, its surrounding context, and multiple relevant helping lines from the buggy source file that are retrieved to augment the surrounding context. This input is then given to the ensemble of checkpoints from the training stage to generate candidate patches. The generated patches are combined, reranked, and validated against the project's test suite to identify a plausible patch that compiles and passes the test cases.
\Cref{fig:arch} illustrates the overall structure of \approach{}. The following sections provide detailed explanations of each component.

\subsection{Data extraction and preprocessing}
\label{subsec:extraction-preprocessing}

To fine-tune a model for automated program repair (APR), we need a large-scale dataset of buggy and fixed code pairs across multiple programming languages. To this end, we follow CoCoNuT \citep{lutellierCoCoNuTCombiningContextaware2020} to extract data from the commit histories of various open-source software repositories. Using real commit history allows the model to access rich, developer-validated examples from multiple programming languages, enhancing its ability to generalize and effectively repair code in various scenarios.

We identify bug-fixing commits by checking for specific keywords such as "bug," "fix," and "patch" in their commit messages \citep{mockusIdentifyingReasonsSoftware2000}. From these commits, we collect pairs of buggy and fixed source code files. We then extract buggy hunks, their buggy context, and the corresponding fixed hunks from the diffs of the buggy and fixed files.

There are different options for selecting the buggy context \citep{chenSequenceRSequencetoSequenceLearning2019, nashidEmbeddingContextCode2023, sintahaKatanaDualSlicing2023}, each with varying levels of effectiveness \citep{yangWhereWereRepair2021}. The buggy context is important as it offers insights into how the broader code structure interacts with the bug. We opt for the function surrounding the buggy hunk as the context because it is easier to obtain in the multilingual setting, is usually short, and carries relevant semantic information about the bug. It is worth noting that our choice of context in the fine-tuning stage is independent of the context used in the inference stage, allowing us to use different context options at either stage.

After extracting the data, we preprocess it following the steps outlined by T5APR \citep{gharibiT5APREmpoweringAutomated2024} to ensure consistency and high-quality instances \citep{raffelExploringLimitsTransfer2020}. In summary, these steps include:
\begin{itemize}
  \item Removing comments from both buggy and fixed hunks, while leaving the context unchanged to preserve semantic information.
  \item Deduplicating instances that share identical buggy hunk, context, and fixed hunk.
  \item Discarding instances with identical buggy and fixed hunks after normalizing whitespace characters.
  \item Removing instances with an empty fixed hunk.
  \item Filtering instances where the token length of the buggy or fixed hunk exceeds the model's input or output limit based on tokenization results.
\end{itemize}

\subsection{Fine-tuning}
\label{subsec:fine-tuning}

We use CodeT5 as our base model, an open-source language model designed to work with code and natural language. CodeT5 adopts the same encoder-decoder transformer architecture of T5 \citep{raffelExploringLimitsTransfer2020} but is pre-trained on a large-scale dataset of code snippets paired with natural language descriptions from different programming languages. The authors of CodeT5 trained the model using the CodeSearchNet dataset \citep{husainCodeSearchNetChallengeEvaluating2020} along with additional data collected from BigQuery. This combined dataset is sourced from public code repositories and contains code snippets from eight programming languages: Ruby, JavaScript, Go, Python, Java, PHP, C, and C\#.

CodeT5's selection is motivated by its effectiveness in code understanding and generation tasks \citep{wangCodeT5IdentifierawareUnified2021} and its adoption in existing APR tools \citep{gharibiT5APREmpoweringAutomated2024, wangRAPGenRetrievalAugmentedPatch2023}.
Furthermore, CodeT5 offers a small version with only 60 million parameters, making it one of the most lightweight encoder-decoder models pre-trained on code. This smaller size allows for better computational efficiency while maintaining a comparable performance to larger versions.

The fine-tuning process updates the parameters of the pre-trained model to improve its ability to handle multilingual APR by minimizing a cross-entropy loss function that measures the discrepancy between the model's predictions and the ground-truth target fixes. We encode the preprocessed collected instances as the model input in the following format:
\begin{verbatim}
language_prefix buggy_hunk : buggy_context
\end{verbatim}
where \verb|language_prefix| is a language-specific prefix added to the beginning of each example to distinguish the programming language (e.g., \verb|Python| or \verb|Java|). The use of a prefix in the encoded input is common in works that use T5-based models \citep{berabiTFixLearningFix2021,raffelExploringLimitsTransfer2020,wangCodeT5IdentifierawareUnified2021,gharibiT5APREmpoweringAutomated2024}. The fixed hunks serve as labels for the model during the fine-tuning process. For each instance, we concatenate all the code lines in the input and label using whitespace to form a single line.

This input is then tokenized using a pre-trained subword byte-level byte-pair-encoding (BPE) \citep{sennrichNeuralMachineTranslation2016} tokenizer from the CodeT5 model and truncated if necessary to ensure that its size remains within the model's maximum input limit. Since we filter buggy hunks with token lengths that exceed the model's input limit in the preprocessing step, any truncation only affects the context part of each instance.

After tokenizing and preparing the instances of each programming language, we concatenate data from all the languages into one dataset. We fine-tune the models on batches that contain samples from all programming languages while using a prefix to identify each language. By starting with a multilingual base model and fine-tuning it with combined data, we enable multilingual learning and allow the model to learn from bug patterns and fixes across different programming languages. This strategy is particularly effective for handling bugs common across all languages \citep{berabiTFixLearningFix2021}.

Due to the diversity of software bugs and their fixes, a single model with optimal parameters may not generalize well \citep{lutellierCoCoNuTCombiningContextaware2020}. Many APR tools use an ensemble of models in one way or another to improve the performance of patch generation for various types of bugs that a single model might not capture \citep{jiangCURECodeAwareNeural2021,jiangKNODDomainKnowledge2023,lutellierCoCoNuTCombiningContextaware2020,wangRAPGenRetrievalAugmentedPatch2023,yeSelfAPRSelfsupervisedProgram2023}. This method is effective at fixing more bugs but is computationally expensive, as it requires training multiple specialized models with varying inputs, hyperparameters, and sometimes architectures.

In contrast, we use the checkpoint ensemble strategy since it has shown to be a cost-effective method that also yields good results \citep{chenCheckpointEnsemblesEnsemble2017, gharibiT5APREmpoweringAutomated2024, huangSnapshotEnsemblesTrain2017}. This approach involves assembling multiple checkpoints captured at distinct intervals during the fine-tuning process, leveraging the model's evolving capabilities as it encounters different subsets of data. The selected checkpoints complement each other to generate patches for various types of bugs and contribute to the quality of patch ranking \citep{gharibiT5APREmpoweringAutomated2024}. Specifically, during fine-tuning over a predefined number of epochs, we save model snapshots at regular intervals of $j$ steps. From this collection, we select an ensemble of $k$ checkpoints for the patch generation step.

\subsection{Context augmentation and patch generation}
\label{subsec:augmentation-generation}

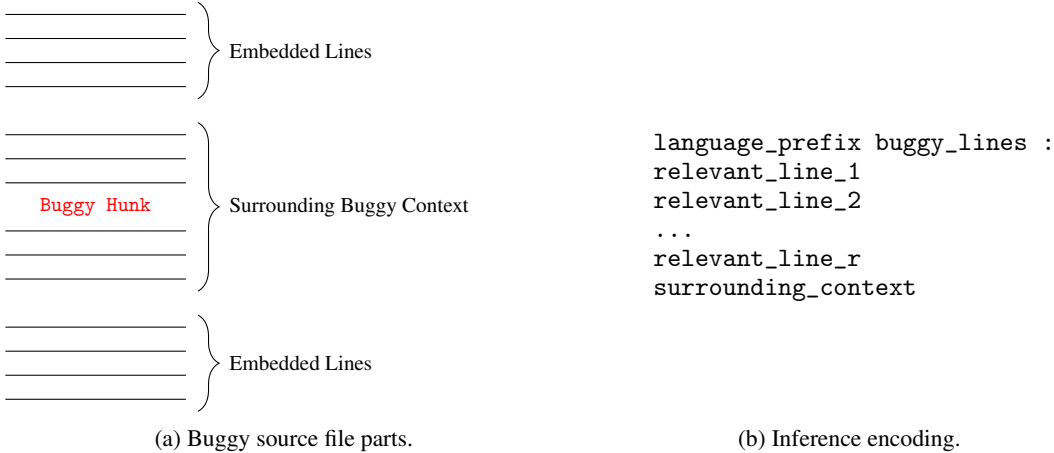
\begin{figure}
  \centering
  \begin{subfigure}[b]{0.45\textwidth}
    \begin{tikzpicture}[scale=0.8, transform shape]

      \foreach \i in {0,...,3} {
          \draw (0, -\i*0.4) -- (3, -\i*0.4);
        }
      \foreach \i in {13,...,16} {
          \draw (0, -\i*0.4) -- (3, -\i*0.4);
        }

      \foreach \i in {5,...,7} {
          \draw (0, -\i*0.4) -- (3, -\i*0.4);
        }
      \foreach \i in {9,...,11} {
          \draw (0, -\i*0.4) -- (3, -\i*0.4);
        }

      \node[red] at (1.5, -8*0.4) {\verb|Buggy Hunk|};

      \node at (3.6, -0.6) [right] {Embedded Lines};
      \node at (3.6, -5.8) [right] {Embedded Lines};
      \node at (3.6, -3.2) [right] {Surrounding Buggy Context};

      \draw[decorate,decoration={brace,amplitude=8pt}] (3.2,0.2) -- (3.2,-1.4);
      \draw[decorate,decoration={brace,amplitude=8pt}] (3.2,-5.0) -- (3.2,-6.6);
      \draw[decorate,decoration={brace,amplitude=8pt}] (3.2,-1.8) -- (3.2,-4.6);

    \end{tikzpicture}
    \caption{Buggy source file parts.}
    \label{fig:source-divide}
  \end{subfigure}
  \begin{subfigure}[b]{0.45\textwidth}
    \begin{verbatim}
      language_prefix buggy_lines :
      relevant_line_1
      relevant_line_2
      ...
      relevant_line_r
      surrounding_context
      \end{verbatim}
    \vspace{1cm}
    \caption{Inference encoding.}
    \label{lst:inference-encoding}
  \end{subfigure}
  \caption{Demonstration of buggy source file parts and input encoding with context augmentation for inference.}
\end{figure}

In this step, we prepare the inference input data from the buggy project to feed into the checkpoints for generating candidate patches. A bug localization approach \citep{gharibiLeveragingTextualProperties2018, vacheretBoostingFaultLocalization2024} is used to identify the project's buggy source files and the location of the bug within them. To construct the inference input, we apply retrieval-augmented generation to augment the buggy context with relevant source lines. This allows the model to access broader contextual information needed to generate the fix. For each buggy hunk, we divide its corresponding source file into three parts: The buggy hunk itself, the surrounding context of the buggy hunk, and the remaining content of the source file, as shown in \cref{fig:source-divide}.

The lines outside the surrounding context are embedded using a vector embedding model. Older search-based APR tools, such as SimFix \citep{jiangShapingProgramRepair2018}, have shown that code snippets similar to the buggy hunk in the project can provide useful ingredients, including donor identifiers or fix patterns for the bug fix. Although embedding lines from the entire project is possible, we limit our analysis scope to the lines in the buggy source file to be more efficient, as it still contains a significant percentage of useful ingredients \citep{barrPlasticSurgeryHypothesis2014,xiaPlasticSurgeryHypothesis2023}. We make sure to deduplicate the embedded lines, remove lines that are empty or only contain punctuations, and exclude the buggy hunk if an identical line exists elsewhere in the file to avoid retrieving the same buggy hunk.

We use dense retrieval \citep{reimersSentenceBERTSentenceEmbeddings2019} to retrieve relevant lines from the source file based on their similarity to the buggy hunk content. Specifically, we embed the buggy hunk using the same embedding model, compute the cosine similarity between the buggy hunk vector and embedded line vectors from the source file, and retrieve up to $r$ most similar lines above a specific similarity threshold. These constraints help reduce noise in the input and prevent exhausting the model's input size. If the buggy hunk is empty, we skip this step.

\Cref{lst:inference-encoding} shows how we use the collected data parts to build the input for patch generation of each hunk. First, we add the language prefix and the buggy hunk. Then, after the \verb|:| separator, we add relevant retrieved lines and finally include the surrounding context. If the buggy hunk is inside a function, the surrounding context is the entire function. In languages that support defining functions inside another function, such as JavaScript and Python, we use the outermost function as the context to account for function closure, where variables from the outer function are used within the inner function. However, if the buggy hunk is outside a function, the surrounding context is the three immediate lines before and after the buggy hunk, including the buggy hunk itself. Adding surrounding lines for buggy lines outside a function is especially important in cases where the fix is purely generative. Without immediate lines, the model would have nothing to work with, as the buggy line is empty and lacks function context.

This input is then tokenized, truncated if necessary, and fed to the checkpoints to generate a ranked list of candidate patches using beam search. We use beam search with a specific beam size $t$ on each checkpoint, generating $t$ best patches from each checkpoint based on the maximum likelihood estimation score of each output sequence. In total, we obtain $k \times t$ patches via the ensemble of $k$ checkpoints for each buggy hunk.

We combine patches for each hunk from different checkpoints to create a single ranked list of candidate patches for each hunk. To this end, we merge all the patches, normalize their whitespace characters, and sort them by their checkpoint rank, breaking ties by the sequence probability score given by the checkpoint. Patches identical to the source hunk are removed, and finally, we remove duplicates and keep the highest-ranked patch from each group of duplicates. We also make sure that if the hunk is not empty, a deletion patch is added at the top of the candidate list. The output is a list of candidate patches for each buggy hunk.

\subsection{Patch validation}
\label{subsec:patch-validation}

Test suite-based APR relies on the test suite to determine program correctness. In this step, we validate the candidate patches generated earlier by applying them to the buggy source code, compiling the patched code, and running the developer-provided test suite. We make sure to omit flaky tests from this process, if any, by running them repeatedly multiple times beforehand. This process discards patches that do not compile or fail to pass the test cases in the project's test suite.
The patched program is considered valid if it passes all the test cases that the original buggy project could pass and successfully passes the triggering test cases that previously failed on the buggy version of the project. The resulting patches that pass the validation process are referred to as plausible patches.

For single-hunk bugs, the ranked list of patches is readily available, allowing us to proceed with validating patches from the top until we find a plausible one that compiles and passes all the test cases.
For multi-hunk bugs, however, the repair process becomes more complex, as the bug may span multiple locations across functions or files, and the combination of patches for all hunks can lead to a huge search space \citep{sahaHarnessingEvolutionMultiHunk2019}. If we generate $t$ patches for each hunk of a bug with $h$ hunks, the total number of candidate patch combinations will be $t^h$. To illustrate with concrete numbers, by generating 100 patches for just 3 hunks, we would need to validate 1,000,000 candidate patches. Inspired by T5APR \citep{gharibiT5APREmpoweringAutomated2024} and ITER \citep{yeITERIterativeNeural2024}, we try to reduce the number of validated patches to about at most $h \times t$, which in this example would be 300 patches. This is fewer than the number of candidate patches some tools validate for a single-hunk bug (\cref{tab:related-work}).

\begin{figure}
  \centering
  \includegraphics{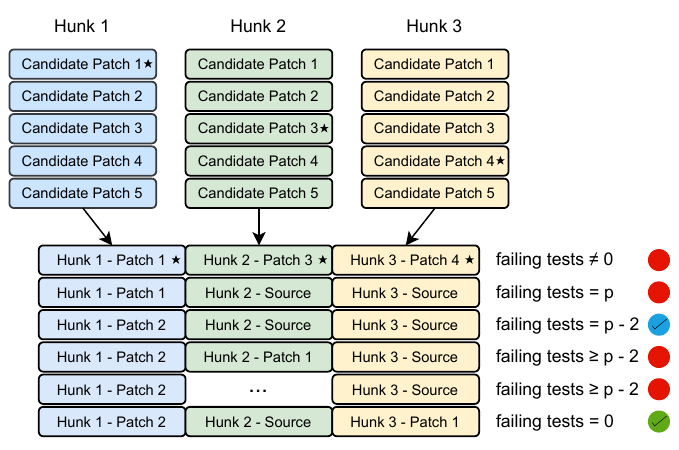}
  \caption{Multi-hunk patch validation process.}
  \label{fig:multihunk-validation}
\end{figure}

\Cref{fig:multihunk-validation} illustrates our multi-hunk validation process. Initially, we attempt to find a plausible patch for all hunks simultaneously. This strategy targets multi-hunk bugs that require the same change across all hunks \citep{madeiralLargescaleStudyHumancloned2021}. We identify generated identical candidate patches for all hunks (starred patches). These patches are then sorted based on the sum of their ranks, followed by their maximum sequence score across all hunks. As a result, we obtain a list of patches that can be uniformly applied to all hunks \citep{gharibiT5APREmpoweringAutomated2024}. If one of these patches passes all the tests, we stop validation and report the plausible patch. Otherwise, we proceed to validate patches for each hunk separately and search for partial patches. Partial patches are patches that do not pass all the tests but pass some of the remaining failing test cases \citep{yeITERIterativeNeural2024}.

When validating patches for the first hunk, other hunks remain unchanged (i.e., keep their source content). If we find a partial patch, we apply it to the current hunk and proceed to the next hunk. If no partial patches are found and all candidate patches for that hunk are tried, we keep the source content for that hunk, hoping that modifications in subsequent hunks may address the dependencies in the previous one. This strategy is useful in scenarios where hunks depend on each other. For example, one hunk might define a variable that is used in another. A patch for the second hunk could both define the variable and use it, eliminating the need to modify the first hunk.

This process of validating patches for each hunk continues until we reach a patch that passes all the tests, or we exhaust all the candidate patches for every hunk. Finally, we combine the resulting patches from all hunks to construct a complete patch for the bug. These plausible patches may not necessarily fix the underlying bug, therefore they should undergo further assessment to determine their correctness in fixing the actual bug \citep{qiAnalysisPatchPlausibility2015}.

\section{Experimental setup}
\label{sec:experimental-setup}

\subsection{Research questions}

This paper aims to address the following research questions:

\begin{itemize}
  \item \textbf{RQ1 (Effectiveness and generalizability)}: How does \approach{} perform in comparison to state-of-the-art APR methods in terms of repair effectiveness and generalizability?
  \item \textbf{RQ2 (Efficiency)}: How efficient is \approach{} in terms of patch ranking and validation cost?
  \item \textbf{RQ3 (Component contribution)}: What are the individual contributions of \approach{}'s context augmentation and multi-hunk patch generation to its overall performance?
\end{itemize}

\subsection{Datasets}

We use a large-scale, multilingual dataset for fine-tuning our repair model, and multiple widely-used benchmarks for evaluating repair effectiveness across different programming languages and bug complexities.

\paragraph{Fine-tuning data}
Since our data extraction process aligns with CoCoNuT \citep{lutellierCoCoNuTCombiningContextaware2020}, we use the dataset provided by CoCoNuT on GitHub\footnote{\url{https://github.com/lin-tan/CoCoNut-Artifact/releases/tag/training_data_1.0.0}} to train \approach{}. To collect this data, commits made before the date of the earliest bug in each evaluation benchmark are extracted to avoid overlap between the training and evaluation data. The dataset contains 8,712,502 tuples of buggy and fixed hunks, along with the surrounding function of each buggy hunk, extracted from the commit history of over 80,000 open-source projects. As we follow the same preprocessing procedure as T5APR \citep{gharibiT5APREmpoweringAutomated2024}, we use 2,324,030 instances remaining after preprocessing and input size filtering.

The dataset covers four programming languages: Java, Python, C, and JavaScript. These languages have widespread use in the software industry, covering a broad range of application domains from web applications to system-level programming, and benefit from a rich availability of training and evaluation datasets. This extensive data enables the model to learn common coding patterns, bug types, and effective fixes better.

Several other works have also used this data \citep{gharibiT5APREmpoweringAutomated2024, jiangCURECodeAwareNeural2021, jiangKNODDomainKnowledge2023, yeNeuralProgramRepair2022, yuanCIRCLEContinualRepair2022}. However, some tools such as Recoder \citep{zhuSyntaxguidedEditDecoder2021} and Tare \citep{zhuTareTypeAwareNeural2023} collect newer data and remove benchmark data from their datasets. In learning-based APR studies, the data used to train the model is considered a key component of the approach, leading to variations across different methods.

\paragraph{Bug benchmarks}
We evaluate the performance of \approach{} using the following benchmarks: Defects4J (Java) \citep{justDefects4JDatabaseExisting2014}, QuixBugs (Java and Python) \citep{linQuixBugsMultilingualProgram2017}, Codeflaws (C) \citep{tanCodeflawsProgrammingCompetition2017}, BugAID (JavaScript) \citep{hanamDiscoveringBugPatterns2016}, BugsInPy (Python) \citep{widyasariBugsInPyDatabaseExisting2020}, and RunBugRun (JavaScript) \citep{prennerRunBugRunExecutableDataset2023}.

These benchmarks cover multiple programming languages and include different types of bugs from real-world software and coding challenges. \citep{sobreiraDissectionBugDataset2018,yeComprehensiveStudyAutomatic2021}.
Defects4J is a curated collection of bugs from 17 well-known open-source Java projects. Following prior work \citep{gharibiT5APREmpoweringAutomated2024, jiangKNODDomainKnowledge2023, mengTemplatebasedNeuralProgram2023, wangRAPGenRetrievalAugmentedPatch2023, weiCopilotingCopilotsFusing2023, xiaLessTrainingMore2022, yeNeuralProgramRepair2022, zhangGammaRevisitingTemplateBased2023, zhuTareTypeAwareNeural2023}, we divide Defects4J into two versions: Defects4J~v1.2 and Defects4J~v2.0. Defects4J~v1.2 contains 395 bugs, and Defects4J~v2.0 contains 444 bugs only available in v2.0.
QuixBugs includes 40 bugs derived from Quixey Challenge problems, consisting of small classic algorithms implemented in single-file programs in both Java and Python.
Codeflaws provides a large-scale collection of 3,903 bugs in C extracted from submissions to Codeforces programming competitions, where each program is a single file.
The BugAID benchmark consists of 12 examples of common JavaScript bug patterns, as described in \citet{hanamDiscoveringBugPatterns2016}.
BugsInPy is a database of real-world bugs from 17 open-source Python projects across different domains.
Lastly, RunBugRun is a multilingual dataset of bugs in nine programming languages, collected from various sources of programming contest submissions. The dataset includes predefined training, validation, and test splits. We use the JavaScript test split from its revised version (v2) for our evaluation.

\begin{table}
  \caption{Statistics of evaluation benchmarks.}
  \centering
  \begin{tabular}{llrrrrr}
    \toprule
    Benchmark      & Language   & Bugs  & Removed & Remained & Single-hunks & Multi-hunks \\
    \midrule
    Defects4J~v1.2 & Java       & 395   & 4       & 391      & 185          & 206         \\
    Defects4J~v2.0 & Java       & 444   & 0       & 444      & 191          & 253         \\
    QuixBugs       & Java       & 40    & 0       & 40       & 40           & 0           \\
    QuixBugs       & Python     & 40    & 0       & 40       & 40           & 0           \\
    Codeflaws      & C          & 3,903 & 8       & 3,895    & 3,552        & 343         \\
    BugAID         & JavaScript & 12    & 0       & 12       & 12           & 0           \\
    BugsInPy       & Python     & 501   & 0       & 501      & 241          & 260         \\
    RunBugRun~v2   & JavaScript & 178   & 0       & 178      & 129          & 49          \\
    \midrule
    Total          &            & 5,513 & 12      & 5,501    & 4,390        & 1,111       \\
    \bottomrule
  \end{tabular}
  \label{tab:benchmarks}
\end{table}

\Cref{tab:benchmarks} provides detailed statistics for each benchmark. It shows the total number of bugs, the number of bugs excluded from consideration because they are either duplicates of other bugs or have no actual changes between their buggy and fixed versions, the remaining bugs used for evaluation, and the number of single and multi-hunk bugs in each benchmark. The classification of single and multi-hunk bugs presented here is based on the data extracted from developer-written patches. While some multi-hunk bugs could be fixed by APR tools with single-location modifications, we classify them as multi-hunk if the corresponding ground-truth patch contains changes at multiple locations.

\subsection{Implementation details and parameters}

\paragraph{Implementation}
We build \approach{} on top of T5APR's codebase in Python to benefit from its efficiency and performance. We use Hugging Face Transformers \citep{wolfTransformersStateoftheArtNatural2020} with PyTorch \citep{paszkePyTorchImperativeStyle2019} backend for fine-tuning and inference. Data processing is handled using Hugging Face Datasets \citep{lhoestDatasetsCommunityLibrary2021} and the Pandas library \citep{wesDataStructuresStatistical2010}. We use Tree-sitter\footnote{\url{https://tree-sitter.github.io/tree-sitter/}} parsers and Pygments\footnote{\url{https://pygments.org/}} lexers to parse and extract relevant data from source files. To parse diffs between buggy and fixed source code files, we use Unidiff\footnote{\url{https://github.com/matiasb/python-unidiff}}.
Additionally, we use the Sentence Transformers library \citep{reimersSentenceBERTSentenceEmbeddings2019} to compute text embeddings. These embeddings are stored and retrieved using Chroma\footnote{\url{https://www.trychroma.com/}}, a vector embedding database optimized for efficient similarity-based search.

For the base model, we use CodeT5-small\footnote{\url{https://huggingface.co/Salesforce/codet5-small}} \citep{wangCodeT5IdentifierawareUnified2021} that has a total of 60M parameters and is pre-trained with an identifier-aware denoising pre-training objective for 100 epochs. We set the maximum input length to 512 and the output length to 256 tokens, the same as what CodeT5 uses in its pre-training and some of its downstream experiments. For the embedding model, we use all-MiniLM-L6-v2\footnote{\url{https://huggingface.co/sentence-transformers/all-MiniLM-L6-v2}}, which maps the input to a 384-dimensional dense vector space.

We reuse the hyperparameters from T5APR to fine-tune our model since they have shown good results, with one difference being that we fine-tune our model for two epochs. Therefore, we use the AdamW optimizer \citep{loshchilovDecoupledWeightDecay2018} with the following hyperparameters: the train batch size is set to 8, the training epochs to 2, the learning rate to $1e-4$, and the learning rate scheduler type to constant. We set $k=5$ and save five checkpoints at every 20\% step of the second epoch during fine-tuning. We also use mixed-precision with FP16 for faster fine-tuning.
For final inference and patch generation, we empirically set the number of retrieved augmented lines ($r$) to five and the cosine similarity threshold to 0.5. This decision is based on preliminary testing on a small set of bugs, as exhaustively evaluating different values across the benchmarks would be time-consuming. This setting also helps avoid saturating the model's input length.
We set the beam size ($t$) to 100 to generate 100 patches from each checkpoint.

We do not claim that the selected models or hyperparameters represent the best possible choices. Rather, we use relatively small models and reasonable hyperparameter settings to demonstrate the effectiveness of our proposed contributions. We acknowledge that adopting larger models \citep{wangCodeT5IdentifierawareUnified2021}, increasing the beam size \citep{tufanoLearningMeaningfulCode2019}, and further tuning the hyperparameters could lead to improved overall performance, but our goal is to highlight the impact of the methodological contributions of our approach.

\paragraph{Infrastructure}
We fine-tune our model on a server equipped with a 4-core Intel Xeon Platinum 8259CL CPU, 16 GB of RAM, and an NVIDIA T4 GPU with 16 GB of VRAM. For evaluation, we use a different system that has a 6-core Intel Core i7-8750H CPU, 16 GB of RAM, and an NVIDIA GeForce GTX 1060 GPU with 6 GB of VRAM.

\subsection{Patch assessment}

Patches that successfully compile and pass the project's test suite are considered plausible, but they may only address the available test cases and fail to generalize to the program's intended behavior \citep{qiAnalysisPatchPlausibility2015}. This issue, known as the test suite overfitting problem, arises when the test suite is weak and does not cover all cases \citep{smithCureWorseDisease2015}. Therefore, it is essential to compare the plausible patches with the developer's patch to assess whether they correctly fix the bug. We consider a patch correct if it meets any of the following criteria \citep{liuEfficiencyTestSuite2020}:

\begin{itemize}
  \item It is identical to the patch provided by the developer.
  \item It matches the patches generated by existing methods that have been reviewed by the community in open-source repositories.
  \item It is semantically equivalent to the developer's patch, meaning it achieves the same functionality and purpose, even if the implementation details differ.
\end{itemize}

In practice, one author verified whether the patches were identical to those created by developers or other existing methods. For patches requiring semantic equivalence checks, the author consulted another author in cases of uncertainty.
To limit potential mistakes in this process, we have made all generated patches publicly available for evaluation and review.\footnote{\url{https://github.com/h4iku/MultiMend/tree/main/results}}

\subsection{Analysis procedure}

We compare \approach{} against recent state-of-the-art APR tools evaluated on our selected benchmarks and report their results under perfect fault localization setting. Perfect fault localization is the preferred way to evaluate APR approaches, as it provides tools with the exact location of bugs and eliminates performance differences caused by varying fault localization techniques. \citep{liuYouCannotFix2019,liuEfficiencyTestSuite2020}.

We select 18 tools for comparison: SequenceR \citep{chenSequenceRSequencetoSequenceLearning2019}, TBar \citep{liuTBarRevisitingTemplatebased2019}, DLFix \citep{liDLFixContextbasedCode2020}, CoCoNuT \citep{lutellierCoCoNuTCombiningContextaware2020}, CURE \citep{jiangCURECodeAwareNeural2021}, Recoder \citep{zhuSyntaxguidedEditDecoder2021}, RewardRepair \citep{yeNeuralProgramRepair2022}, CIRCLE \citep{yuanCIRCLEContinualRepair2022}, \citet{prennerCanOpenAICodex2022}, \citet{sobaniaAnalysisAutomaticBug2023}, KNOD \citep{jiangKNODDomainKnowledge2023}, AlphaRepair \citep{xiaLessTrainingMore2022}, RAP-Gen \citep{wangRAPGenRetrievalAugmentedPatch2023}, Repilot \citep{weiCopilotingCopilotsFusing2023}, Tare \citep{zhuTareTypeAwareNeural2023}, TENURE \citep{mengTemplatebasedNeuralProgram2023}, Gamma \citep{zhangGammaRevisitingTemplateBased2023}, and T5APR \citep{gharibiT5APREmpoweringAutomated2024}.

From this list, TBar is a traditional template-based approach, while AlphaRepair, Repilot, and Gamma use large language models for cloze-style infilling repair, with the remaining tools being learning-based. Additionally, AlphaRepair, TENURE, and Gamma enhance their performance through crafted fix templates. \citet{prennerCanOpenAICodex2022} and \citet{sobaniaAnalysisAutomaticBug2023} use OpenAI's Codex and ChatGPT, respectively, under similar settings as the other tools, with ChatGPT evaluated by generating patches through independent requests, without utilizing its conversational capabilities.

We obtain the results of other tools from their papers, their code repositories with updated results, or other papers that evaluate them under our ideal settings \citep{liuEfficiencyTestSuite2020,zhongStandUp4NPRStandardizingSetUp2023}.
We consider only the first generated plausible patch by \approach{} for each bug that successfully compiles and passes the test suite, following previous works \citep{durieuxEmpiricalReviewJava2019,liuTBarRevisitingTemplatebased2019,lutellierCoCoNuTCombiningContextaware2020}.
Furthermore, we tried our best to report the first plausible correctness result for other tools. However, since some tools do not explicitly specify whether they consider only the first generated plausible patch or all generated patches, we report their results as provided.

To compute the patch ranking information for other methods, we obtain the list of generated candidate patches from their code repositories, where available, and extracted the ranking positions based on the reported plausible and correct patches for each bug in each benchmark.

\section{Results and discussion}
\label{sec:results}

\subsection{RQ1: Effectiveness and generalizability}

\begin{table}
  \caption{The number of fixed bugs and comparison with state-of-the-art approaches. Results are presented as \textit{correct/plausible (identical)}, where values in parentheses are bugs with identical patches to the developer's. A dash (-) indicates data unavailability or the tool does not support the programming language or benchmark. The highest number of correctly fixed bugs for each benchmark is highlighted in bold.}
  \centering
  \scriptsize
  \tabcolsep=0.1cm
  \begin{tabular}{lcccccccc}
    \toprule
                                                                  & Defects4J~v1.2                          & Defects4J~v2.0       & QuixBugs            & QuixBugs            & Codeflaws                    & BugAID           & BugsInPy             & RunBugRun~v2          \\
    Tool                                                          & 391 bugs (J)                            & 444 bugs (J)         & 40 bugs (J)         & 40 bugs (P)         & 3,895 bugs (C)               & 12 bugs (JS)     & 501 bugs (P)         & 178 bugs (JS)         \\
    \midrule
    SequenceR \citep{chenSequenceRSequencetoSequenceLearning2019} & 12/19 (10)                              & -                    & 15/16 (15)          & -                   & -                            & -                & -                    & -                     \\
    TBar \citep{liuTBarRevisitingTemplatebased2019}               & 53/84 (36)                              & -                    & -                   & -                   & -                            & -                & -                    & -                     \\
    DLFix \citep{liDLFixContextbasedCode2020}                     & 39/68 (34)                              & -                    & -                   & -                   & -                            & -                & -                    & -                     \\
    CoCoNuT \citep{lutellierCoCoNuTCombiningContextaware2020}     & 44/85 (26)                              & 21/31 (16)           & 13/20 (13)          & 19/21 (15)          & 423/716 (255)                & 3/- (3)          & -                    & -                     \\
    CURE \citep{jiangCURECodeAwareNeural2021}                     & 57/104 (37)                             & 19/- (9)             & 25/34 (20)          & -                   & -                            & -                & -                    & -                     \\
    Recoder \citep{zhuSyntaxguidedEditDecoder2021}                & 64/69 (46)                              & -                    & 17/17 (-)           & -                   & -                            & -                & -                    & -                     \\
    RewardRepair \citep{yeNeuralProgramRepair2022}                & 45/- (38)                               & 45/- (42)            & 20/- (20)           & -                   & -                            & -                & -                    & -                     \\
    CIRCLE \citep{yuanCIRCLEContinualRepair2022}                  & 64/182 (49)                             & -                    & -                   & -                   & -                            & 5/- (5)          & -                    & -                     \\
    \citet{prennerCanOpenAICodex2022}                             & -                                       & -                    & 14/- (-)            & 23/- (-)            & -                            & -                & -                    & -                     \\
    \citet{sobaniaAnalysisAutomaticBug2023}                       & -                                       & -                    & -                   & 19/- (13)           & -                            & -                & -                    & -                     \\
    KNOD \citep{jiangKNODDomainKnowledge2023}                     & 71/85 (49)                              & 50/82 (27)           & 25/30 (19)          & -                   & -                            & -                & -                    & -                     \\
    AlphaRepair \citep{xiaLessTrainingMore2022}                   & 74/109 (44)                             & 36/50 (30)           & \textbf{28/30 (24)} & 27/32 (23)          & -                            & -                & -                    & -                     \\
    RAP-Gen \citep{wangRAPGenRetrievalAugmentedPatch2023}         & 59/- (54)                               & 47/- (38)            & -                   & -                   & -                            & -                & -                    & -                     \\
    Repilot \citep{weiCopilotingCopilotsFusing2023}               & 66/- (51)                               & 50/- (40)            & -                   & -                   & -                            & -                & -                    & -                     \\
    Tare \citep{zhuTareTypeAwareNeural2023}                       & 75/126 (44)                             & 55/107 (22)          & 27/27 (19)          & -                   & -                            & -                & -                    & -                     \\
    TENURE  \citep{mengTemplatebasedNeuralProgram2023}            & 77/- (48)                               & 50/- (36)            & -                   & -                   & -                            & -                & -                    & -                     \\
    Gamma \citep{zhangGammaRevisitingTemplateBased2023}           & \textbf{80/99 (59)}                     & 45/- (-)             & 22/- (-)            & -                   & -                            & -                & -                    & -                     \\
    T5APR \citep{gharibiT5APREmpoweringAutomated2024}             & 67/94 (46)                              & 56/103 (35)          & 25/26 (18)          & 29/30 (24)          & 1,764/2,359 (1,259)          & 5/- (5)          & -                    & -                     \\
    \midrule
    \approach{}                                                   & 79/127 (53)                             & \textbf{70/132 (44)} & \textbf{28/30 (18)} & \textbf{31/32 (25)} & \textbf{1,864/2,460 (1,309)} & \textbf{6/- (6)} & \textbf{49/225 (30)} & \textbf{100/109 (60)} \\
    \midrule
    Total                                                         & \multicolumn{8}{c}{2,227/3,121 (1,545)}                                                                                                                                                                     \\
    \bottomrule
  \end{tabular}
  \label{tab:main-results}
\end{table}

\paragraph{Overall results}
\Cref{tab:main-results} presents the performance of \approach{} in comparison with several state-of-the-art automated program repair (APR) tools across multiple benchmarks, including Defects4J~v1.2 and v2.0 (Java), QuixBugs (Java \& Python), Codeflaws (C), BugAID (JavaScript), BugsInPy (Python), and RunBugRun~v2 (JavaScript). The results are reported as the number of correct/plausible patched bugs, with the number of patches identical to the developer fixes provided in parentheses. Correctly fixed bugs in this table are from the first plausible patches generated by each APR technique. The table shows the total number of considered bugs, with the abbreviated language name for each benchmark below its name. The BugAID benchmark does not contain tests to validate the patches; therefore, we only report the number of bugs with identical patches and cannot provide the count of plausible ones.

Overall, \approach{} repairs 2,227 bugs correctly, with 1,545 of them identical to the developer's fix. \approach{} outperforms or matches the compared tools in terms of correct fixes across all benchmarks, except for Defects4J~v1.2, where Gamma correctly fixes one additional bug. The number of patches identical to developer fixes is also competitive. Specifically, on Defects4J~v2.0, QuixBugs (Python), Codeflaws, and BugAID, \approach{} generates more identical patches, indicating its ability to generate developer-like fixes.

On the Defects4J~v1.2 benchmark, \approach{} achieves 79 correct fixes out of 127 plausible patches, with 53 of these patches identical to the developer's patch. This places \approach{} among the top-performing tools and second only to Gamma, which produces 80 correct fixes. Furthermore, \approach{} demonstrates strong performance on the Defects4J~v2.0 benchmark, achieving 70 correct fixes out of 132 plausible patches (44 identical), outperforming all other approaches. It is worth mentioning that Gamma, AlphaRepair, and TENURE use templates to improve their patch generation performance, a technique that could be integrated into our tool in the future to further enhance its effectiveness.

In the QuixBugs benchmark, a smaller dataset of 40 Java and Python programs, \approach{} maintains competitive results. For QuixBugs (Java), \approach{} achieves 28 correct fixes, outperforming all other tools except AlphaRepair, which also fixes the same amount of bugs but with more identical patches. On QuixBugs (Python), \approach{} achieves 31 correct fixes, surpassing all other approaches.

On the Codeflaws dataset, \approach{} produces 1,864 correct fixes, 1,309 of which are identical to the developer's fixes. This performance establishes \approach{} as the most effective tool for this large-scale dataset, significantly outperforming other tools. Similarly, on the BugAID benchmark, \approach{} successfully repairs six bugs correctly. While BugAID is a relatively small benchmark, the results further shows \approach{}'s consistent performance.

For BugsInPy and RunBugRun, the compared tools are not evaluated on these benchmarks. However, \approach{}'s performance on them provides a strong baseline for future comparisons. On BugsInPy, \approach{} correctly fixes 49 bugs, with 30 identical patches to the developer's. The large gap between plausible and correct fixes in this benchmark is because the provided relevant tests for most bugs are not very comprehensive. In contrast, RunBugRun includes many test cases for most of its bugs, resulting in a high correctness ratio, where 100 out of 109 plausible fixes are correct.

\approach{} demonstrates generalizability by effectively repairing bugs in statically and dynamically typed programming languages, and producing high-quality patches, as evidenced by a high proportion of identical fixes. Its robust performance across diverse benchmarks, including large-scale benchmarks (Codeflaws), smaller curated benchmarks (QuixBugs, BugAID, and RunBugRun), and project-based benchmarks (Defects4J and BugsInpy), highlights its adaptability to varying codebases and bug distributions, making it a promising tool for real-world program repair scenarios.

\begin{figure}
  \captionsetup[subfigure]{size=scriptsize, labelformat=empty}
  \centering
  \begin{subfigure}[b]{.3\textwidth}
    \centering
    \includegraphics{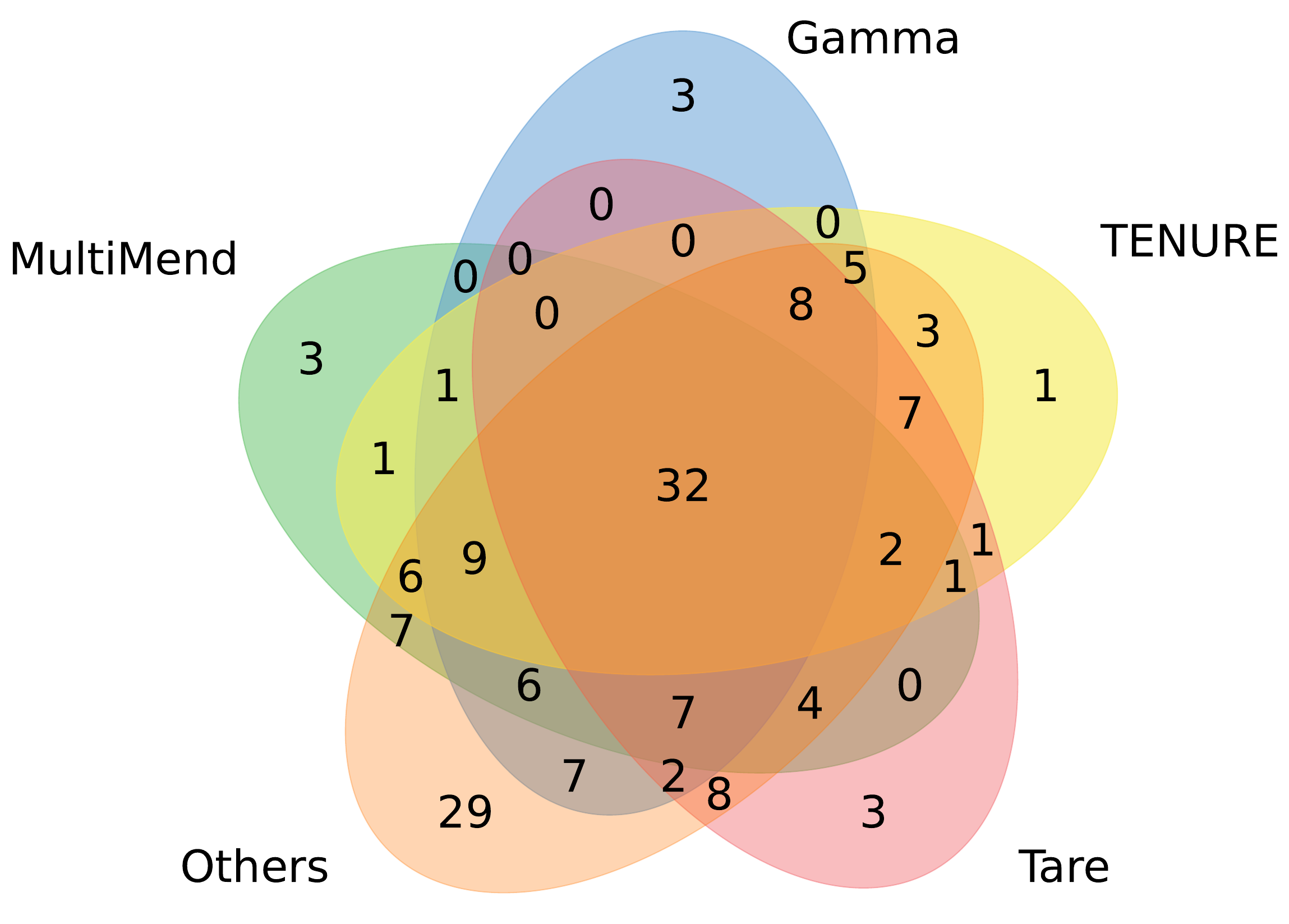}
    \caption{\small Defects4J~v1.2 (Java)}
  \end{subfigure}%
  \hfill
  \begin{subfigure}[b]{.3\textwidth}
    \centering
    \includegraphics{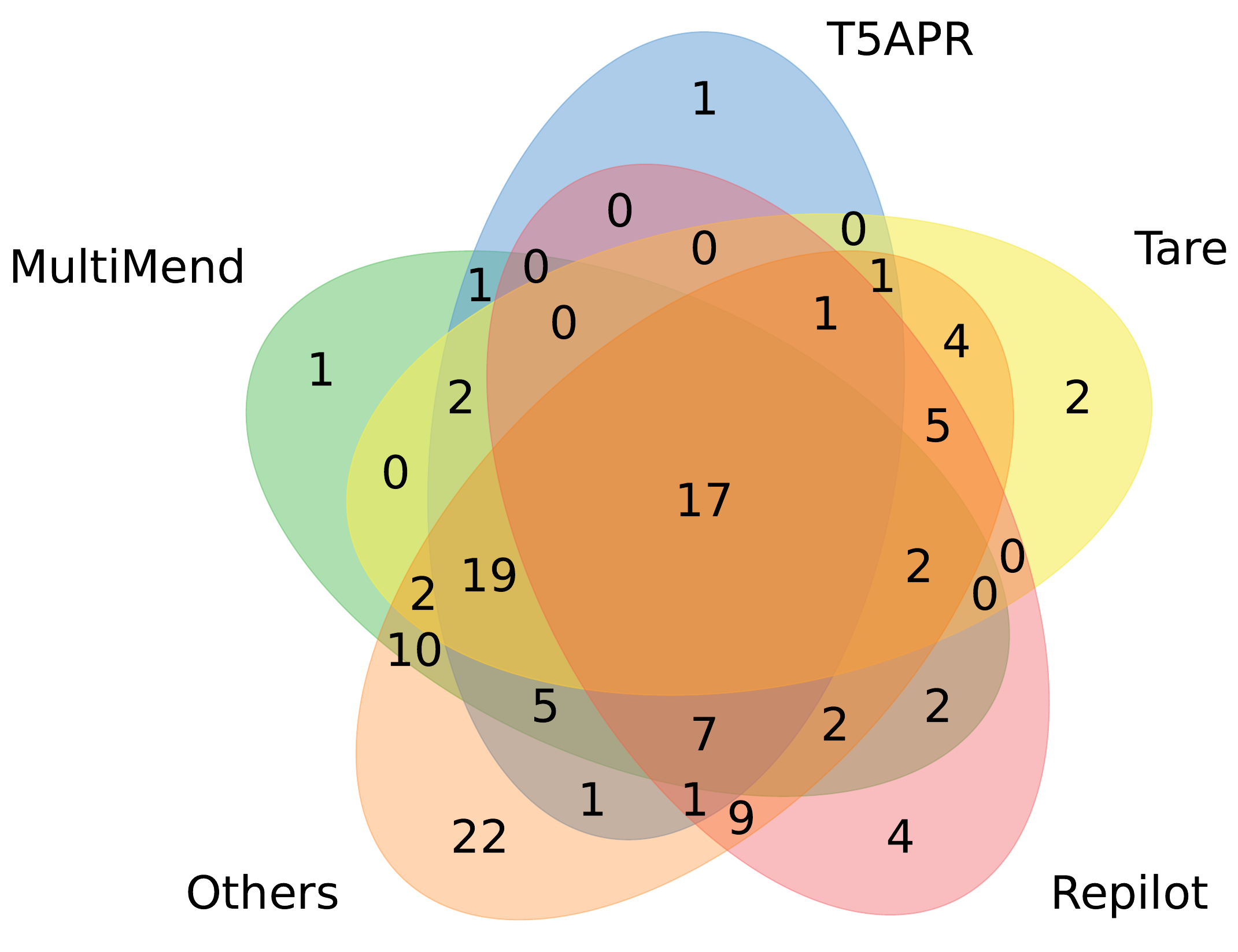}
    \caption{\small Defects4J~v2.0 (Java)}
  \end{subfigure}%
  \hfill
  \begin{subfigure}[b]{.3\textwidth}
    \centering
    \includegraphics{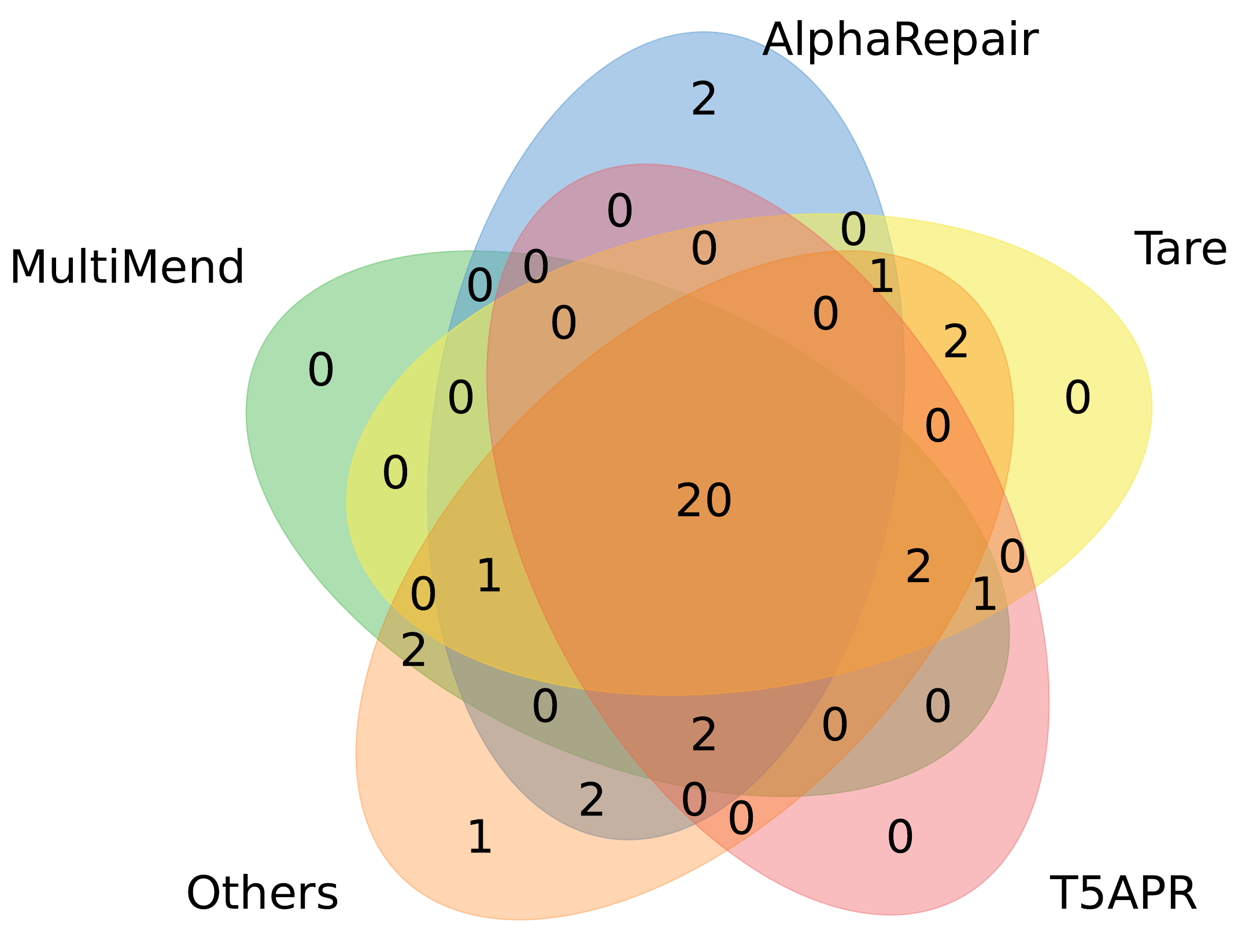}
    \caption{\small QuixBugs (Java)}
  \end{subfigure}
  \begin{subfigure}[b]{.3\textwidth}
    \centering
    \includegraphics{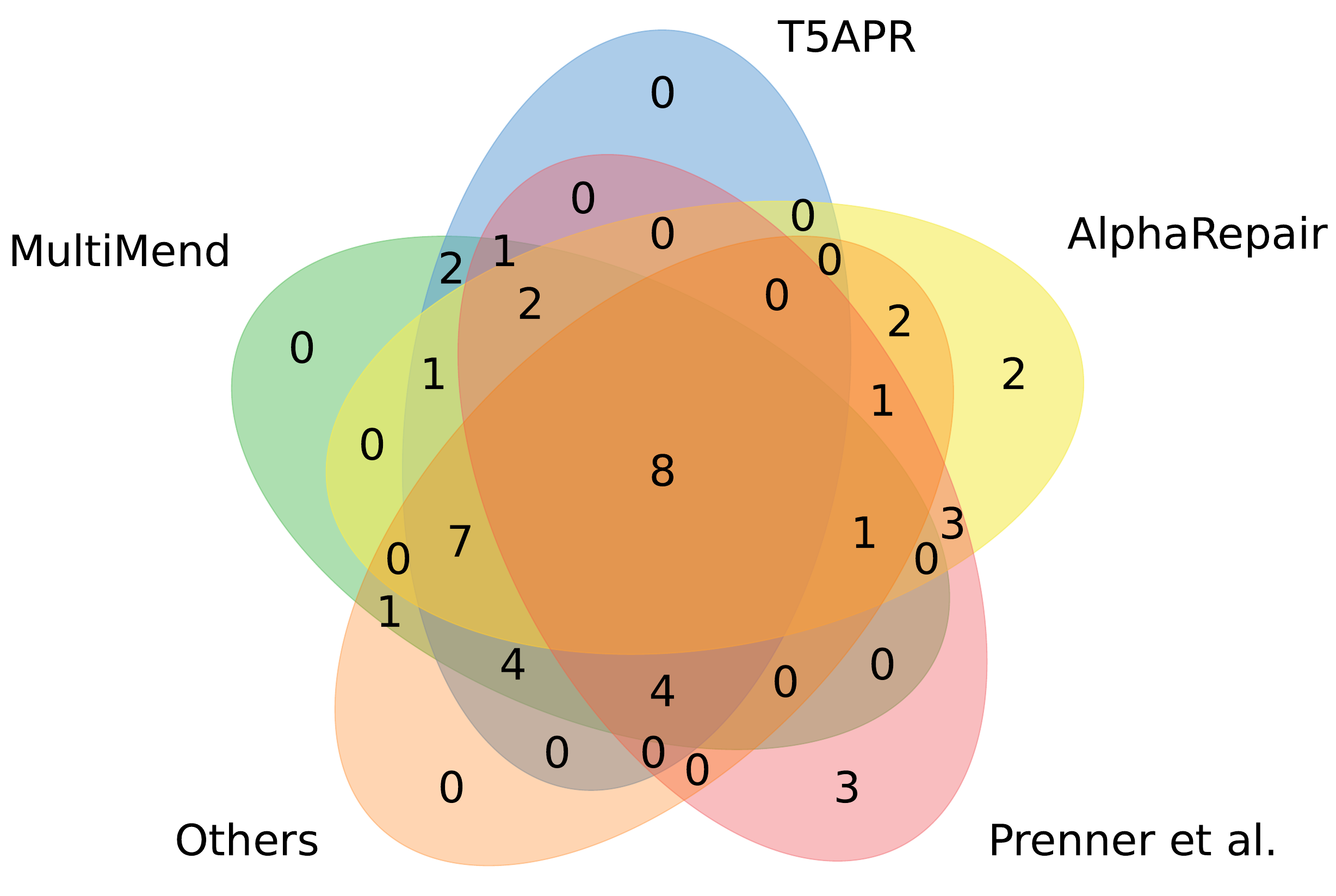}
    \caption{\small QuixBugs (Python)}
  \end{subfigure}%
  \hfill
  \begin{subfigure}[b]{.3\textwidth}
    \centering
    \includegraphics{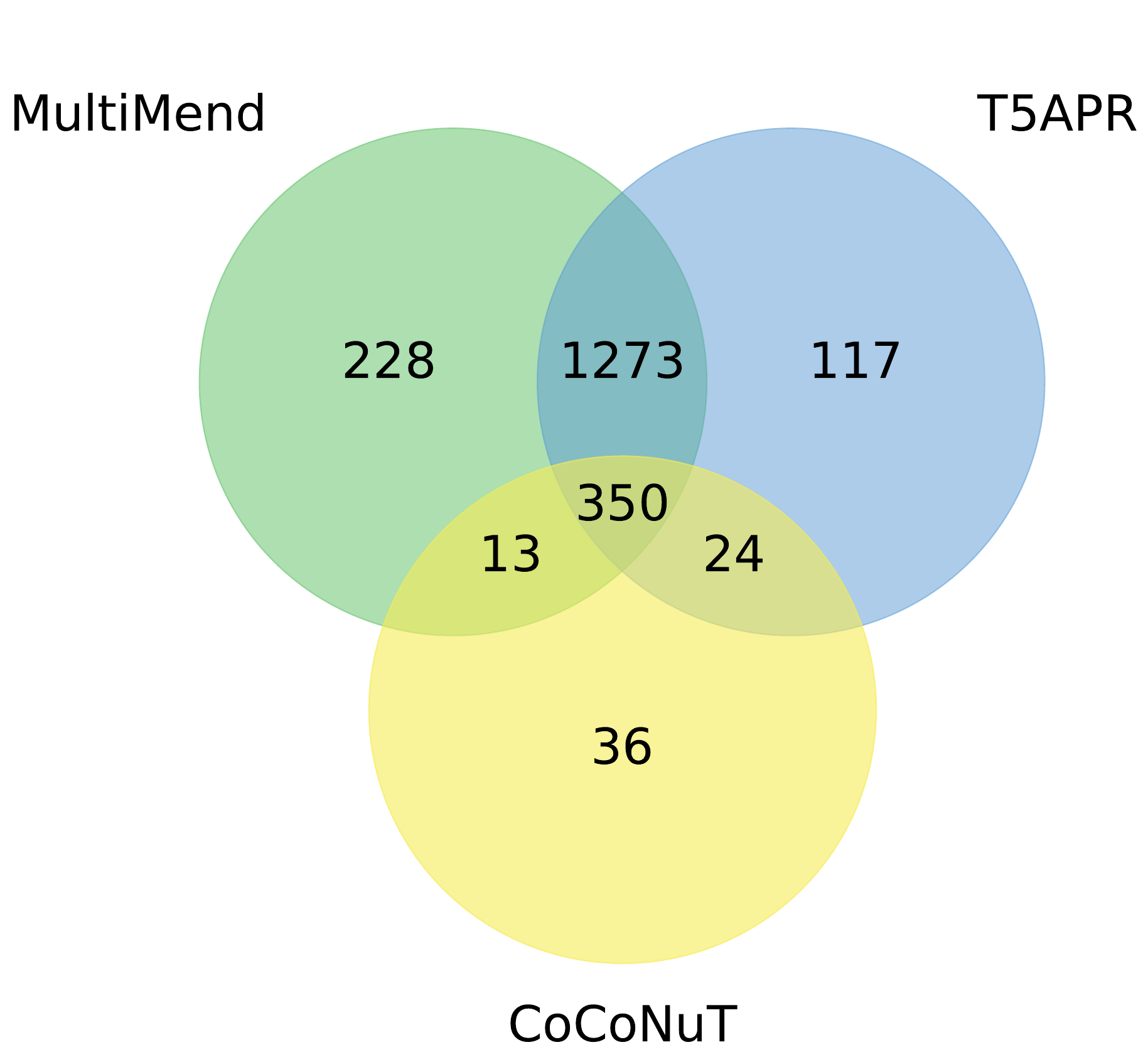}
    \caption{\small Codeflaws (C)}
  \end{subfigure}%
  \hfill
  \begin{subfigure}[b]{.3\textwidth}
    \centering
    \includegraphics{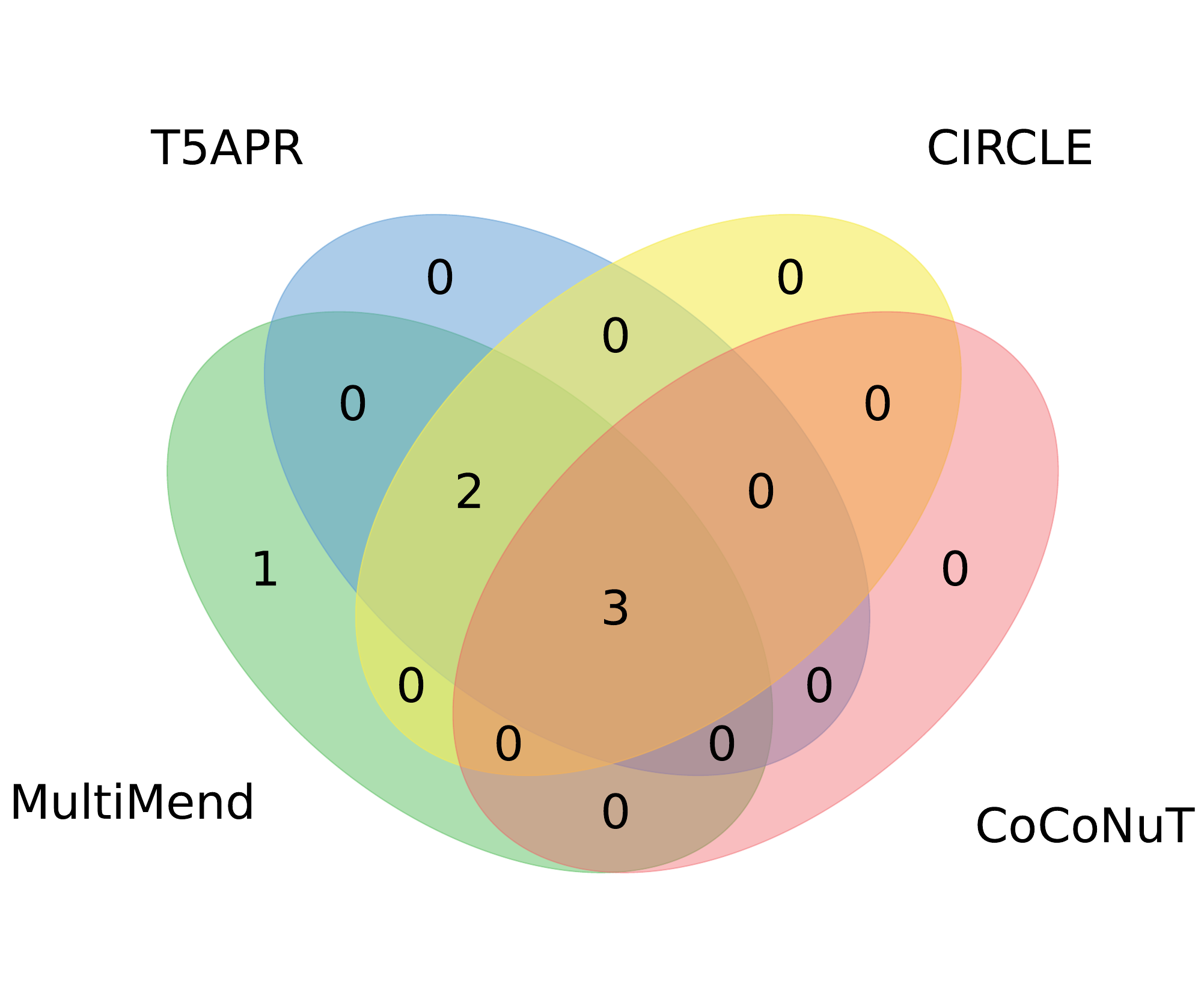}
    \caption{\small BugAID (JavaScript)}
  \end{subfigure}
  \caption{Number of unique and overlapped bug fixes of \approach{} and other tools.}
  \label{fig:overlap-fixes}
\end{figure}

\paragraph{Unique and overlapping bug fixes}
\Cref{fig:overlap-fixes} demonstrates the unique and overlapping bugs repaired by different approaches across the evaluated benchmarks. For benchmarks used for evaluation by more than four tools, we select the top three tools for that benchmark and group the bugs fixed by the remaining tools under the "Others" category.

Across the shown benchmarks, \approach{} fixes 233 bugs that other tools do not. Due to the high number of proposed approaches, most have substantial overlap in the bugs they repair. Defects4J has the highest number of evaluated tools, resulting in the Others group having the most number of unique fixes. In QuixBugs, being a small benchmark, \approach{} does not fix any previously unfixed bugs but achieves the highest number of simultaneous fixes overall.

The differences in unique bug repairs reveal that no single method dominates across all benchmarks. However, this shows an opportunity that combining multiple techniques could enhance overall effectiveness. Future work could explore integrating these techniques or developing ensemble methods to maximize repair coverage \citep{zhangNeuralProgramRepair2023, zhongPracticalProgramRepair2024}.

\begin{figure}
  \centering
  \begin{minted}[
      mathescape,
      autogobble,
      fontsize=\small,
      frame=lines,
      framesep=2mm,
      ]{diff}
         if (objectType != null) {
           // Is this a normal property access, or are we trying to override
           // an existing property?
      -    boolean isOverride = t.inGlobalScope() &&
      +    boolean isOverride = parent.getJSDocInfo() != null &&
           parent.getType() == Token.ASSIGN &&
           parent.getFirstChild() == getprop;
    \end{minted}
  \caption{Fix generated for the Closure~71 bug from Defects4J~v1.2.}
  \label{lst:fix-closure-71}
\end{figure}

\begin{figure}
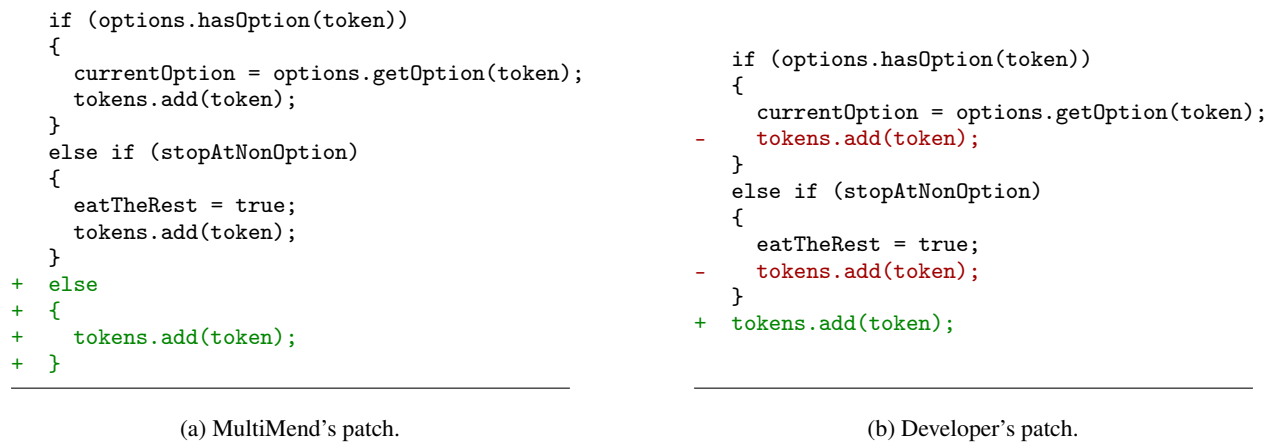

  \centering
  \begin{subfigure}[b]{.45\textwidth}
    \begin{minted}[
      mathescape,
      autogobble,
      fontsize=\small,
      frame=lines,
      framesep=2mm,
      ]{diff}
         if (options.hasOption(token))
         {
           currentOption = options.getOption(token);
           tokens.add(token);
         }
         else if (stopAtNonOption)
         {
           eatTheRest = true;
           tokens.add(token);
         }
      +  else
      +  {
      +    tokens.add(token);
      +  }
    \end{minted}
    \caption{\approach{}'s patch.}
    \label{lst:fix-cli-19-tool}
  \end{subfigure}%
  \hfill
  \begin{subfigure}[b]{.45\textwidth}
    \begin{minted}[
      mathescape,
      autogobble,
      fontsize=\small,
      frame=lines,
      framesep=7.18mm,
      ]{diff}
         if (options.hasOption(token))
         {
           currentOption = options.getOption(token);
      -    tokens.add(token);
         }
         else if (stopAtNonOption)
         {
           eatTheRest = true;
      -    tokens.add(token);
         }
      +  tokens.add(token);
    \end{minted}
    \caption{Developer's patch.}
    \label{lst:fix-cli-19-dev}
  \end{subfigure}
  \caption{Fix for Cli~19 bug from Defects4J~v2.0.}
  \label{lst:fix-cli-19}
\end{figure}

\begin{figure}
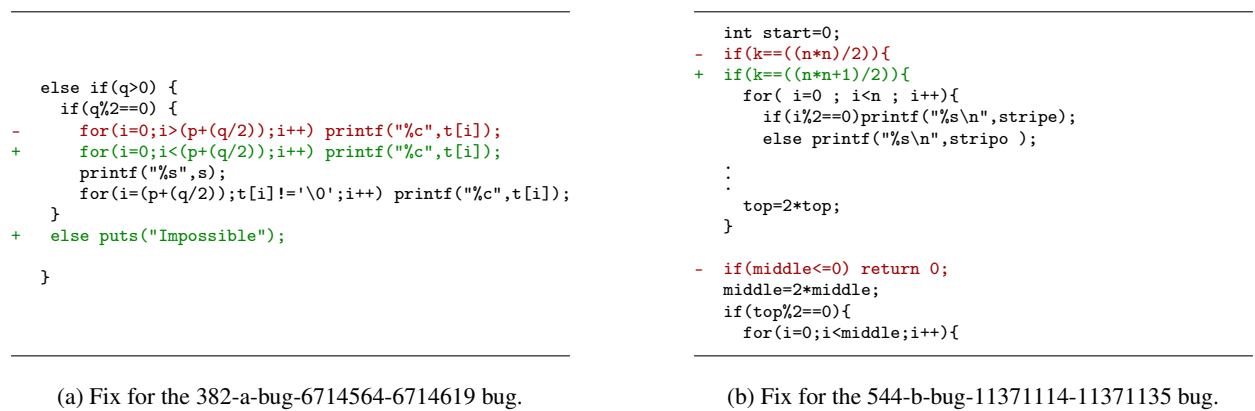

  \centering
  \begin{subfigure}[b]{.45\textwidth}
    \begin{minted}[
      mathescape,
      autogobble,
      fontsize=\scriptsize,
      frame=lines,
      framesep=9.36mm,
      ]{diff}
         else if(q>0) {
           if(q%2==0) {
      -      for(i=0;i>(p+(q/2));i++) printf("%c",t[i]);
      +      for(i=0;i<(p+(q/2));i++) printf("%c",t[i]);
             printf("%s",s);
             for(i=(p+(q/2));t[i]!='\0';i++) printf("%c",t[i]);
          }
      +   else puts("Impossible");
      
         }
    \end{minted}
    \caption{Fix for the 382-a-bug-6714564-6714619 bug.}
    \label{lst:fix-codeflaws-1}
  \end{subfigure}%
  \hfill
  \begin{subfigure}[b]{.45\textwidth}
    \begin{minted}[
      mathescape,
      autogobble,
      fontsize=\scriptsize,
      frame=lines,
      framesep=2mm,
      escapeinside=\#\%
      ]{diff}
         int start=0;
      -  if(k==((n*n)/2)){
      +  if(k==((n*n+1)/2)){
           for( i=0 ; i<n ; i++){
             if(i%2==0)printf("%s\n",stripe);
             else printf("%s\n",stripo );
         #\vdots%
           top=2*top;
         }
  
      -  if(middle<=0) return 0;
         middle=2*middle;
         if(top%2==0){
           for(i=0;i<middle;i++){
    \end{minted}
    \caption{Fix for the 544-b-bug-11371114-11371135 bug.}
    \label{lst:fix-codeflaws-2}
  \end{subfigure}
  \caption{Multi-hunk fixes generated for Codeflaws bugs.}
  \label{lst:fix-codeflaws}
\end{figure}

\Cref{lst:fix-closure-71,lst:fix-cli-19,lst:fix-codeflaws} showcase some of the unique bugs fixed by \approach{}. \Cref{lst:fix-closure-71} shows the fix for the Closure~71 bug from the Defects4J~v1.2 benchmark. This bug has a large surrounding context, and its fix is generated only with the augmented context. Adding contradictory conditions based on \verb|t.inGlobalScope()| to the input leads the model to question the correctness of the logic and focus on exploring alternative conditional values related to the \verb|isOverride| variable to come up with the fix. \verb|parent.getJSDocInfo()| appears in an \verb|if| block, where its condition depends on \verb|isOverride| and initializes a variable called \verb|overridingInfo|. This patch is identical to the developer's patch.

\Cref{lst:fix-cli-19} presents the patch generated for the multi-hunk bug Cli~19 from Defects4J~v2.0 and its developer's ground-truth patch. Although the developer's patch modifies multiple locations to fix the bug, \approach{}, following its multi-hunk strategy, decides to keep the code in the first two hunks and adds a final hunk to generate a semantically equivalent patch.

\Cref{lst:fix-codeflaws} demonstrates unique multi-hunk patches generated for two bugs in the Codeflaws benchmark. \Cref{lst:fix-codeflaws-1} shows the multi-hunk fix generated for the 382-a-bug-6714564-6714619 bug. The fix is nearly identical to the developer's patch, except that the developer's fix uses \verb|printf| instead of \verb|puts| in the second hunk. This patch highlights the \approach{}'s capability to combine different partial patches into a full patch to fix a bug.
Similarly, \Cref{lst:fix-codeflaws-2} shows the multi-hunk fix generated for the 544-b-bug-11371114-11371135 bug, which matches the developer's patch. In this bug, the hunks are further apart from each other.

\subsection{RQ2: Efficiency}

\begin{figure}
  \centering
  \includegraphics{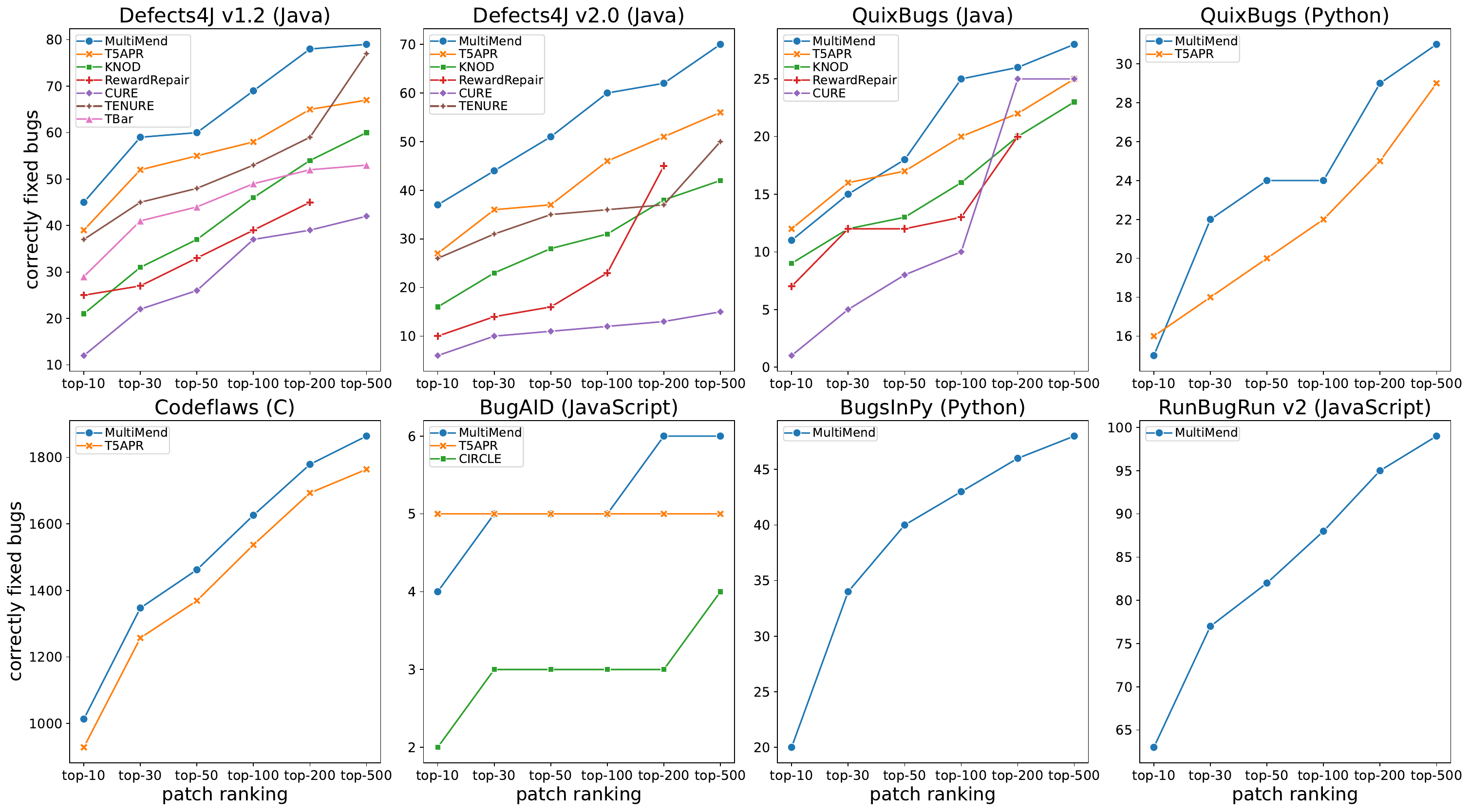}
  \caption{Patch ranking information of correctly fixed bugs.}
  \label{fig:ranking-info}
\end{figure}

\paragraph{Patch ranking}
To understand the patch generation efficiency of \approach{}, we analyze the ranking position of the correct patches. The ranking position of correct patches for each approach is extracted from the generated patch list of tools provided in their software repositories. We then compute the number of correctly fixed bugs at various ranking thresholds. The results are presented in \cref{fig:ranking-info}.

On Defects4J~v1.2, \approach{} outperforms all competing approaches across all ranking thresholds. As the threshold increases, \approach{} maintains a significant lead over other approaches, with only TENURE coming close at top-500. A similar trend is observed on the Defects4J~v2.0 benchmark, where \approach{} shows a steady improvement through the thresholds and outperforms all other techniques. T5APR and TENURE show competitive performance but remain consistently behind \approach{}. Some tools that do not achieve their full result in the top-500 threshold (e.g., KNOD on Defects4J~v1.2) fix the rest of their bugs at higher thresholds due to using larger beam sizes and generating more candidate patches overall.

On the QuixBugs (Java \& Python) and BugAID benchmarks, \approach{} achieves the best performance at higher thresholds. T5APR follows closely, showing better performance at the lower thresholds but lagging as the threshold increases. A contributing factor to \approach{}'s weaker performance at the start is its difference in ranking candidate patches. \approach{} always ranks the empty patch at the beginning, whereas T5APR only does this if the empty patch is manually added to the candidate patch list, leaving already generated empty patches in their original positions. On Codeflaws, \approach{} maintains a steady lead over its competitor in all thresholds. BugsInPy and RunBugRun also show a clear upward trend as the threshold increases.

Across all the benchmark datasets, \approach{} demonstrates competitive effectiveness in ranking correct patches higher. Its performance advantage is particularly notable on larger datasets such as Codeflaws and Defects4J, where it achieves substantial gains over other approaches.
Overall, \approach{} ranks 343 of its correct fixes at the top of the candidate patch list, with 329 of them identical to the developer's fixes.

\begin{figure}
  \centering
  \includegraphics{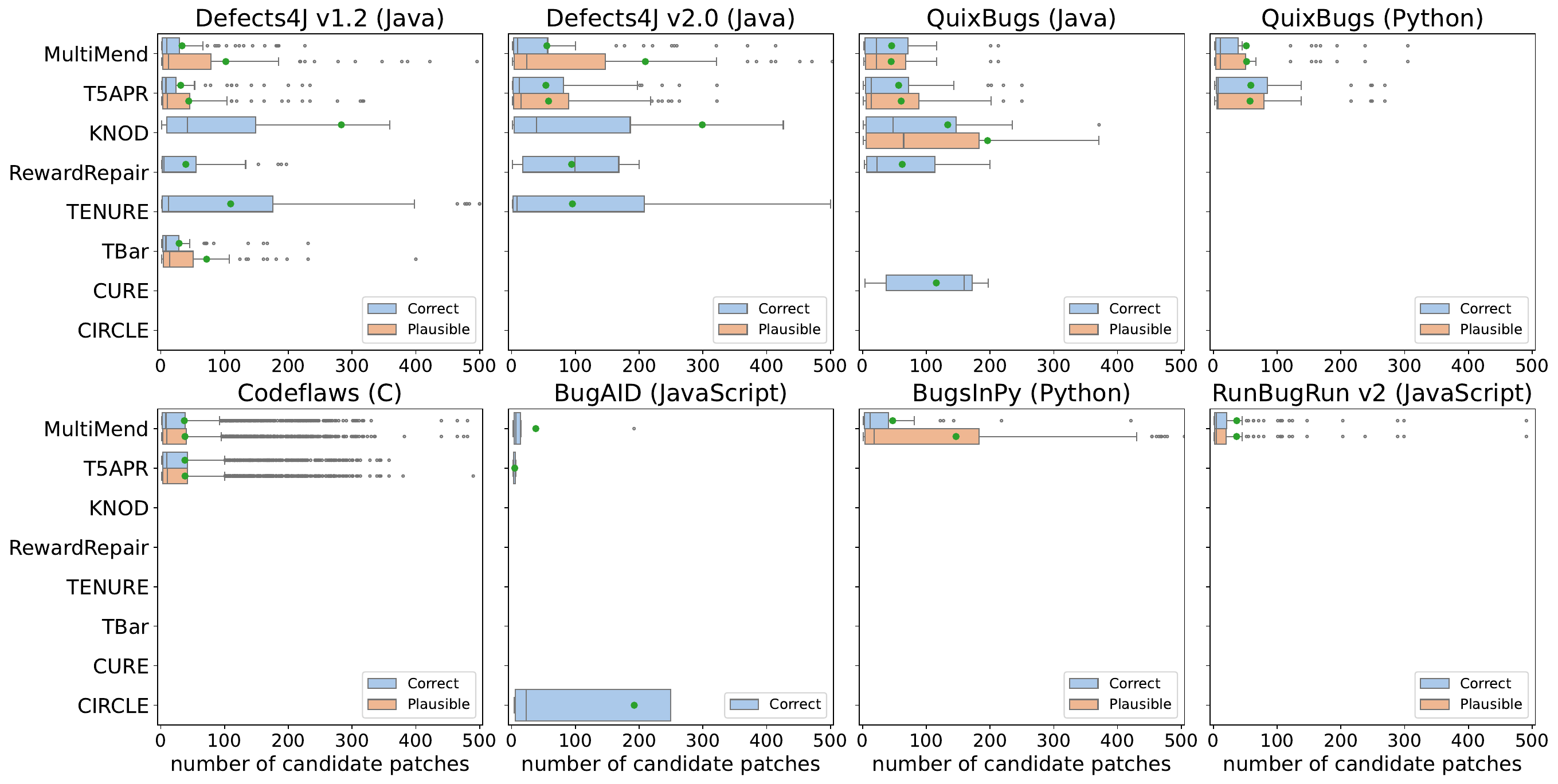}
  \caption{Number of candidate patches before correct or plausible.}
  \label{fig:npc-boxplot}
\end{figure}

\paragraph{Number of candidate patches}
To further analyze the efficiency of \approach{}, \cref{fig:npc-boxplot} shows the distribution of the number of patch candidates (NPC) validated by APR tools across multiple benchmarks until a correct or plausible patch is found \citep{liuEfficiencyTestSuite2020, qiUsingAutomatedProgram2013}. This analysis considers only bugs for which a plausible or correct patch exists.
The box plots summarize the statistics for each tool, distinguishing between correct patches (blue, upper plot) and plausible patches (orange, lower plot). Green dots indicate the mean values and the horizontal axis represents the number of candidate patches, capped at 500 for visualization clarity. Plausible patch data was unavailable for some approaches, so in these cases, only the correct patch statistics are shown.

Across all benchmarks, \approach{} demonstrates a compact range of validated patches compared to most other tools, with relatively small interquartile ranges for both correct and plausible patches. This indicates that \approach{} achieves correct and plausible repairs more consistently within a smaller number of candidate patches. The only competitors on most benchmarks are T5APR and TBar, which occasionally work better with fewer validated candidate patches. This improvement, particularly for plausible patches, can be attributed to the fact that \approach{} targets more multi-hunk bugs, which leads to validating more patches overall, especially when dealing with a higher number of hunks.
In contrast, other tools, such as KNOD, RewardRepair, and TENURE, exhibit significantly wider ranges, reflecting greater variability in patch validation and validating many patches.

\begin{table}
  \caption{Statistics for validation until finding a correct or plausible patch. Times are in HH:MM:SS.fff format.}
  \centering
  \small
  \tabcolsep=0.12cm
  \begin{tabular}{lccccccccc}
    \toprule
    \multirow{2}{*}{Benchmark}                  &
    \multicolumn{4}{c}{Time to correct}         &
    \multicolumn{4}{c}{Validated until correct} &
    \multicolumn{1}{c}{Timeouts}                                                                                                                        \\
    \cmidrule(r){2-5} \cmidrule(r){6-9} \cmidrule(r){10-10}
                                                & {min}        & {max}        & {median}     & {mean}       & {min} & {max} & {median} & {mean} & {max} \\
    \midrule
    Defects4J~v1.2 {\scriptsize(Java)}          & 00:00:07.149 & 01:11:21.255 & 00:01:17.682 & 00:05:55.509 & 1     & 226   & 9.0      & 33.29  & 2     \\
    Defects4J~v2.0 {\scriptsize(Java)}          & 00:00:04.130 & 00:14:00.626 & 00:00:39.467 & 00:02:30.519 & 1     & 414   & 9.5      & 55.17  & 0     \\
    QuixBugs {\scriptsize(Java)}                & 00:00:08.190 & 00:25:48.788 & 00:01:46.332 & 00:03:34.840 & 2     & 213   & 22.0     & 45.71  & 9     \\
    QuixBugs {\scriptsize(Python)}              & 00:00:00.617 & 00:21:04.746 & 00:00:03.427 & 00:01:21.155 & 2     & 305   & 11.0     & 51.65  & 21    \\
    Codeflaws {\scriptsize(C)}                  & 00:00:00.192 & 01:49:13.592 & 00:00:02.033 & 00:00:46.660 & 1     & 481   & 8.0      & 36.87  & 108   \\
    BugsInPy {\scriptsize(Python)}              & 00:00:00.563 & 00:15:36.899 & 00:00:10.535 & 00:01:09.817 & 1     & 517   & 12.0     & 47.33  & 0     \\
    RunBugRun~v2 {\scriptsize(JavaScript)}      & 00:00:00.092 & 03:18:38.046 & 00:00:20.228 & 00:04:52.167 & 1     & 564   & 5.0      & 36.67  & 19    \\
    \midrule
    Overall                                     & 00:00:00.092 & 03:18:38.046 & 00:00:03.106 & 00:01:15.085 & 1     & 564   & 8.0      & 37.86  & 108   \\
    \bottomrule
  \end{tabular}

  \bigskip

  \begin{tabular}{lccccccccc}
    \toprule
    \multirow{2}{*}{Benchmark}                    &
    \multicolumn{4}{c}{Time to plausible}         &
    \multicolumn{4}{c}{Validated until plausible} &
    \multicolumn{1}{c}{Timeouts}                                                                                                                          \\
    \cmidrule(r){2-5} \cmidrule(r){6-9} \cmidrule(r){10-10}
                                                  & {min}        & {max}        & {median}     & {mean}       & {min} & {max} & {median} & {mean} & {max} \\
    \midrule
    Defects4J~v1.2 {\scriptsize(Java)}            & 00:00:07.149 & 03:14:58.922 & 00:01:34.844 & 00:10:15.586 & 1     & 2244  & 12.0     & 101.95 & 2     \\
    Defects4J~v2.0 {\scriptsize(Java)}            & 00:00:04.130 & 20:02:32.445 & 00:01:21.767 & 00:26:14.931 & 1     & 4288  & 23.5     & 209.69 & 6     \\
    QuixBugs {\scriptsize(Java)}                  & 00:00:08.190 & 00:31:34.026 & 00:01:46.332 & 00:04:24.440 & 2     & 213   & 22.0     & 44.93  & 28    \\
    QuixBugs {\scriptsize(Python)}                & 00:00:00.617 & 00:21:04.746 & 00:00:03.577 & 00:01:19.280 & 2     & 305   & 11.0     & 52.13  & 21    \\
    Codeflaws {\scriptsize(C)}                    & 00:00:00.183 & 02:44:19.990 & 00:00:02.202 & 00:00:56.341 & 1     & 481   & 9.0      & 37.88  & 161   \\
    BugsInPy {\scriptsize(Python)}                & 00:00:00.266 & 03:59:29.780 & 00:00:25.316 & 00:05:44.636 & 1     & 2097  & 18.0     & 146.67 & 15    \\
    RunBugRun~v2 {\scriptsize(JavaScript)}        & 00:00:00.092 & 03:18:38.046 & 00:00:19.904 & 00:04:39.823 & 1     & 564   & 5.0      & 36.44  & 19    \\
    \midrule
    Overall                                       & 00:00:00.092 & 20:02:32.445 & 00:00:03.728 & 00:02:54.376 & 1     & 4288  & 10.0     & 55.80  & 161   \\
    \bottomrule
  \end{tabular}
  \label{tab:val-cost}
\end{table}

To complement these plots, \cref{tab:val-cost} provides detailed statistics for validation time costs, validated patch counts, and the maximum number of timeouts of \approach{} before reaching a correct or plausible patch.

The time required to find a correct or plausible patch shows significant variation across benchmarks due to differences in tooling, programming language characteristics, type of bugs, and size of projects in each benchmark.
The overall median time to a plausible patch is significantly low at about three seconds, indicating that plausible patches are identified quickly in most cases. However, the mean times reveal the impact of outliers, particularly in the Defects4J~v2.0 benchmark, which has the highest mean time of about 26 minutes. These outliers are primarily caused by bugs with a large number of hunks (e.g., JacksonDatabind 103 with 26 hunks). Additionally, patch validation timeouts, which require waiting for a fixed duration until the validation of the applied patch times out before proceeding, contribute substantially to increasing the overall validation time. This is evident in the Codeflaws benchmark, where it has very low minimum times but also high maximum times with a lot of timeouts. These patterns show opportunities for optimization, such as improved patch ranking mechanisms.

The overall median NPC value until finding a correct or plausible patch is eight and ten patches, respectively, suggesting that relatively few validations are required in most cases. However, as mentioned earlier, certain benchmarks, such as Defects4J~v2.0, contain multi-hunk bugs with many hunks, resulting in outliers with high validation counts of up to 4,288 patches.
For most benchmarks, the mean and median to reach a correct patch are lower than those for a plausible patch, suggesting that correct patches, when present in the candidate list, are identified faster \citep{liuEfficiencyTestSuite2020}. A possible explanation for this observation is that correct patches often involve idiomatic and more natural transformations that align with common programming patterns learned from historical fixes during training, causing them to rank higher in candidate lists. However, as we move down the candidate list, the chance of finding patches that exploit gaps in test coverage and bypass the test suite through contrived modifications increases.

Overall, in our main experiment, we validated 1,434,638 patches across all benchmarks that required about 65 days of execution time. Fine-tuning \approach{}'s model took about 35 hours, and generating candidate patches for each hunk takes five seconds (one second per checkpoint).

\subsection{RQ3: Component contribution}

\begin{table}
  \caption{Comparison of results with and without context augmentation. Results are shown as \textit{correct/plausible}.}
  \centering
  \begin{tabular}{lcc}
    \toprule
    Benchmark                 & Without context augmentation & With context augmentation \\
    \midrule
    Defects4J~v1.2 (Java)     & 79/126                       & 79/127                    \\
    Defects4J~v2.0 (Java)     & 69/126                       & 70/132                    \\
    QuixBugs (Java)           & 27/29                        & 28/30                     \\
    QuixBugs (Python)         & 31/32                        & 31/32                     \\
    Codeflaws (C)             & 1,845/2,460                  & 1,864/2,460               \\
    BugAID (JavaScript)       & 6/-                          & 6/-                       \\
    BugsInPy (Python)         & 47/224                       & 49/225                    \\
    RunBugRun~v2 (JavaScript) & 100/109                      & 100/109                   \\
    \midrule
    Total                     & 2,204/3,112                  & 2,227/3,121               \\
    \bottomrule
  \end{tabular}
  \label{tab:rag-norag}
\end{table}

\Cref{tab:rag-norag} provides a comparative analysis of the effectiveness of input context augmentation using retrieval-augmented generation (RAG) to generate patches. The results show a consistent matching or improvement in both correctness and plausibility when context augmentation is applied, albeit with varying degrees of impact depending on the benchmark. These findings suggest that the benefit of context augmentation on simpler benchmarks or smaller codebases is less pronounced, potentially due to the limited availability of contextual information to aid the repair. For instance, QuixBugs programs are mostly single functions and there is usually nothing more to be retrieved. When aggregating all benchmarks, the use of context augmentation yields 2,227 correct patches and 3,121 plausible patches, compared to 2,204 correct and 3,112 plausible patches without it.

Other than QuixBugs (Java \& Python) BugAID, and RunBugRun where the bugs fixed with context augmentation form a superset of those fixed without it, there are differences in the bugs fixed by each scenario in other benchmarks. Some bugs are fixed with context augmentation but not without it, and vice versa. With context augmentation, a total of 40 unique correct fixes are generated that are not found without context augmentation, while the baseline approach generates 17. Context augmentation also enables exploring a broader patch space and enhances the variety of plausible fixes with 22 unique patches.

The presence of unique fixes in both categories highlights that context augmentation can serve as a complementary enhancement but will not always improve performance. It is important to balance the use of context augmentation to avoid the risk of introducing irrelevant or conflicting information, which could confuse the model and degrade its performance. This observation aligns with the findings of \citet{prennerOutContextHow2024}. Combining patches from both configurations (with and without augmented context) might yield the best outcome. Future work could focus on refining the context selection process or even allowing the model to choose its required context \citep{parasaramFactSelectionProblem2025} for fixing a bug to amplify its benefits while mitigating its limitations.
It is worth noting that we fine-tuned the model using only the buggy lines and their surrounding function. Augmenting the context in the fine-tuned data might have improved robustness to noisy inputs during inference. However, we do not have access to the complete source files of the fine-tuning data, and collecting them again would be expensive.

\begin{figure}
  \centering
  \begin{subfigure}[b]{\textwidth}
    \begin{minted}[
      mathescape,
      autogobble,
      fontsize=\small,
      frame=lines,
      framesep=2mm,
      ]{diff}
         protected XmlSerializerProvider(XmlSerializerProvider src) {
           super(src);
      -    _rootNameLookup = src._rootNameLookup;
      +    _rootNameLookup = new XmlRootNameLookup();
         }
    \end{minted}
    \caption{Patch generated for the JacksonXml~5 bug from Defects4J~v2.0.}
    \label{lst:fix-jacksonxml-5}
  \end{subfigure}%
  \vspace{5mm}
  \begin{subfigure}[b]{\textwidth}
    \begin{minted}[
      mathescape,
      autogobble,
      fontsize=\small,
      frame=lines,
      framesep=2mm,
      ]{diff}
         for (int i = getNumObjectiveFunctions(); i < getArtificialVariableOffset(); i++) {
           final double entry = tableau.getEntry(0, i);
      -    if (Precision.compareTo(entry, 0d, maxUlps) > 0) {
      +    if (Precision.compareTo(entry, 0d, epsilon) > 0) {
             columnsToDrop.add(i);
           }
         }
    \end{minted}
    \caption{Patch generated for the Math~33 bug from Defects4J~v1.2.}
    \label{lst:fix-math-33}
  \end{subfigure}
  \caption{Examples of fixes generated only with augmented context.}
  \label{lst:fix-only-rag}
\end{figure}

Examples of fixes only generated with context augmentation are shown in \Cref{lst:fix-only-rag}. \Cref{lst:fix-jacksonxml-5} shows the patch for the JacksonXml~5 bug from Defects4J~v2.0. In this bug, the identifier \verb|XmlRootNameLookup|, which needed to be used as the repair ingredient, is not present in the surrounding method but is defined as a field in the same class as `\mintinline[breaklines]{text}{protected final XmlRootNameLookup _rootNameLookup;}.' The retrieval mechanism adds this line to the input and the model generates the correct fix, identical to the developer's patch. \Cref{lst:fix-math-33} provides another example, showing the patch for Math~33 from Defects4J~v1.2. This patch is also identical to the developer's patch. In this bug, the identifier \verb|epsilon|, required to fix the bug, is not in the surrounding method. However, the retriever identifies the line `\mintinline[breaklines]{text}{if (Precision.compareTo(entry, 0d, epsilon) < 0)}' in another method of the same file that guides the model in the right direction.

\begin{figure}
  \centering
  \begin{subfigure}[b]{\textwidth}
    \begin{minted}[
      mathescape,
      autogobble,
      fontsize=\small,
      frame=lines,
      framesep=2mm,
      ]{diff}
         public Paint getPaint(double value) {
           double v = Math.max(value, this.lowerBound);
           v = Math.min(v, this.upperBound);
      -    int g = (int) ((value - this.lowerBound) / (this.upperBound
      +    int g = (int) ((v - this.lowerBound) / (this.upperBound
                   - this.lowerBound) * 255.0);
           return new Color(g, g, g);
         }
    \end{minted}
    \caption{Patch without context augmentation.}
    \label{lst:fix-chart-24-norag}
  \end{subfigure}%
  \vspace{5mm}
  \begin{subfigure}[b]{\textwidth}
    \begin{minted}[
      mathescape,
      autogobble,
      fontsize=\small,
      frame=lines,
      framesep=2mm,
      ]{diff}
         public Paint getPaint(double value) {
           double v = Math.max(value, this.lowerBound);
           v = Math.min(v, this.upperBound);
      -    int g = (int) ((value - this.lowerBound) / (this.upperBound
      +    int g = (int) ((int)(value - lowerBound) / (upperBound
                   - this.lowerBound) * 255.0);
           return new Color(g, g, g);
         }    
    \end{minted}
    \caption{Patch with context augmentation.}
    \label{lst:fix-chart-24-rag}
  \end{subfigure}
  \caption{Patches generated for Chart~24 from Defects4J~v1.2.}
  \label{lst:fix-chart-24}
\end{figure}

\Cref{lst:fix-chart-24} is an example where the non-augmented input fixes the bug, but the augmented input fails and finds an incorrect plausible patch first. These patches are generated for the Chart~24 bug from Defects4J~v1.2. Although the correct patch is also generated lower in the candidate list with context augmentation, the overemphasis on type casting in the input causes the plausible but incorrect patch to rank higher and pass the test cases due to limitations in the test suite.

\begin{table}
  \caption{Comparison of single and multi-hunk results. Results are shown as \textit{correct/plausible}.}
  \centering
  \begin{tabular}{lcc}
    \toprule
                              & Single-hunk & Multi-hunk \\
    Benchmark                 & 4,390 bugs  & 1,111 bugs \\
    \midrule
    Defects4J~v1.2 (Java)     & 61/88       & 18/39      \\
    Defects4J~v2.0 (Java)     & 60/96       & 10/36      \\
    QuixBugs (Java)           & 28/30       & 0/0        \\
    QuixBugs (Python)         & 31/32       & 0/0        \\
    Codeflaws (C)             & 1,786/2,346 & 78/114     \\
    BugAID (JavaScript)       & 6/-         & 0/0        \\
    BugsInPy (Python)         & 46/145      & 3/80       \\
    RunBugRun~v2 (JavaScript) & 88/93       & 12/16      \\
    \midrule
    Total                     & 2,106/2,836 & 121/285    \\
    \bottomrule
  \end{tabular}
  \label{tab:single-multi}
\end{table}

\Cref{tab:single-multi} presents the results of \approach{} on single-hunk and multi-hunk bug fixes across various benchmarks. The total number of available single-hunk and multi-hunk bugs is listed under the table header.
\approach{} demonstrates strong performance in single-hunk scenarios, achieving 2,106 correct and 2,836 plausible fixes. Additionally, our multi-hunk strategy produces 121 correct and 285 plausible multi-hunk patches, highlighting \approach{}'s capacity to handle more complex bug-fixing scenarios.

An advantage of our approach to multi-hunk bugs is that there is no limitation on the number of hunks or their locations, as they are treated independently. Furthermore, it does not require a model specifically trained for multi-hunk repair; the same model used for single-hunk repair can be applied to multi-hunk bugs. However, this strategy has a limitation as it relies on failing tests to identify partial fixes and guide the repair process of multi-hunk bugs.
A common observation with multi-hunk bugs in our experiment is that addressing a single hunk may sometimes pass all tests. This occurs because developers do not always write tests that cover all related hunks, and fixing just one hunk might be sufficient to satisfy the existing tests. Multi-hunk repair therefore requires a stronger test suite, both to localize relevant hunks and to drive the repair from partial patches to a complete patch.
Leveraging models with larger input sizes and intelligently grouping multiple hunks from the same function, hunks in close distance, or hunks associated with shared test cases into a single input can help mitigate this limitation \citep{huangEmpiricalStudyFineTuning2023,silvaRepairLLaMAEfficientRepresentations2025}.

Overall, \approach{}'s context augmentation and multi-hunk patch generation mechanisms contribute significantly to its competitive edge.

\subsection{Threats to validity}

In this section, we discuss potential threats to the validity of our experimental results and describe the measures taken to mitigate them.

\paragraph{Internal validity}
One significant threat is the possibility of errors in the manual assessment of patch correctness, a common threat in APR research \citep{tianBestBothWorlds2023}. The correctness of the generated plausible patches is evaluated manually, and lack of domain knowledge about the programs, human errors, or biases during this process could lead to the misclassification of patches as correct or incorrect \citep{yeAutomatedPatchAssessment2021}.
While every effort is made to ensure the correctness of manual assessment, given the scale of the generated patches and the human involvement, errors may still exist. To mitigate this issue, we carefully reviewed the plausible patches, compared them with those produced by existing tools, and made the results publicly available.

Another potential threat arises from faults in the approach's implementation. Bugs or unintended behaviors in the implementation could skew the results, either overstating or understating the tool's effectiveness. To address this, we carefully checked our implementation and verified the outputs at multiple stages to identify and rectify potential issues. Additionally, we make the source code of our tool publicly available to allow for external review and feedback.

Lastly, there is a risk of data leakage due to the use of CodeT5, which is pre-trained on a large corpus of publicly available code. If any portion of the benchmark datasets overlaps with this training corpus, it could affect the tool's performance on those benchmarks. This issue is particularly challenging to address since retraining the model would be extremely costly. Other widely used language models, such as CodeBERT \citep{fengCodeBERTPreTrainedModel2020}, Codex, and ChatGPT, as well as tools built on these models, face a similar challenge \citep{prennerCanOpenAICodex2022, sobaniaAnalysisAutomaticBug2023, xiaLessTrainingMore2022}. However, several factors mitigate this concern. First, any overlapping data in the model's pre-training corpus constitutes only a small fraction of the total data. Additionally, CodeT5 is not specifically trained for APR and its pre-training objectives differ from our task. This issue is also less critical in program repair because the model does not encounter pairs of buggy code snippets and their fixes during pre-training, nor do such pairs include explicit labels indicating whether they are buggy or correct \citep{jiangImpactCodeLanguage2023}. Furthermore, after fine-tuning, we anticipate that any leaked knowledge from CodeT5 will be forgotten. Our fine-tuning data is curated up to the date of the bugs in our evaluation benchmarks, ensuring no overlap between them.

\paragraph{External validity}
The generalizability of our approach may pose a threat to external validity due to a phenomenon known as benchmark overfitting \citep{durieuxEmpiricalReviewJava2019}. While our approach has shown promising results on the selected benchmarks, these benchmarks represent only a subset of real-world bugs. As a result, the proposed approach may not generalize effectively to other types of bugs that are not covered in the tested benchmarks. To mitigate this issue, we evaluated \approach{} on six benchmarks across different programming languages, covering 5,501 bugs. This makes our study one of the most comprehensive evaluations in APR. We still acknowledge that future work should extend testing to additional benchmarks and real-world bugs, such as GitBug-Java \citep{silvaGitBugJavaReproducibleBenchmark2024}, BugsJS \citep{gyimesiBugsJSBenchmarkJavaScript2019}, and BugsCpp \citep{anBugsCHighlyUsable2023}, to better understand the broader applicability and limitations of our tool.

\section{Related work}
\label{sec:related-work}

Automated program repair (APR) has been an active area of research in recent years, with numerous techniques and tools proposed to address the challenging problem of automatically fixing bugs in software to reduce maintenance costs and improve its reliability \citep{gaoProgramRepair2022,huangSurveyAutomatedProgram2023}. Early APR approaches focused on generating patches using search-based, constraint-based, and template-based techniques.

Search-based methods, such as GenProg \citep{legouesGenProgGenericMethod2012} and SimFix \citep{jiangShapingProgramRepair2018}, iteratively explore syntactic code edits to find plausible patches. Constraint-based techniques, including SemFix \citep{nguyenSemFixProgramRepair2013}, Angelix \citep{mechtaevAngelixScalableMultiline2016}, and Maple \citep{nguyenAutomaticProgramRepair2019}, use constraint-solving and formal specifications to synthesize repairs that maintain intended program behavior. Template-based methods apply predefined \citep{liuYouCannotFix2019} or mined \citep{koyuncuFixMinerMiningRelevant2020,liuTBarRevisitingTemplatebased2019} templates to correct specific types of bugs, providing targeted, reusable repair patterns.

More recent work has explored the use of machine learning and deep learning techniques to use the power of data-driven models, utilizing large code repositories and historical bug-fix data to improve the efficiency and effectiveness of program repair \citep{zhangSurveyLearningbasedAutomated2023,zhangSystematicLiteratureReview2024}. Various approaches are proposed in this area. One approach is supervised learning, where models learn to repair code by being trained or fine-tuned on large datasets of buggy and corrected code, allowing them to generalize to similar errors \citep{chenSequenceRSequencetoSequenceLearning2019,jiangCURECodeAwareNeural2021,lutellierCoCoNuTCombiningContextaware2020,yeNeuralProgramRepair2022,dingPatchingTranslationData2021,liDLFixContextbasedCode2020,gharibiT5APREmpoweringAutomated2024,wangRAPGenRetrievalAugmentedPatch2023,jiangKNODDomainKnowledge2023}. In contrast, self-supervised methods use unlabeled data to generate training examples of code errors and corrections, enabling the model to learn from synthetic or inferred examples without requiring extensive bug-fix pairs \citep{allamanisSelfSupervisedBugDetection2021,yasunagaGraphbasedSelfSupervisedProgram2020,yasunagaBreakItFixItUnsupervisedLearning2021,yeSelfAPRSelfsupervisedProgram2023,silvaMUFINImprovingNeural2023}. Conversation-based methods use instruction-tuned large language models (LLMs) with natural language instructions enriched with various ingredients at each stage to guide models through the repair process, allowing them to follow specific commands or repair objectives effectively \citep{cheshkovExploringPotentialConversational2024, kongContrastRepairEnhancingConversationBased2025, xiaAutomatedProgramRepair2024}. Agent-based methods involve agentic workflows and instruct LLMs to collaborate in detecting, analyzing, and fixing bugs. For example, one agent might identify potential bug locations while another proposes and validates fixes \citep{bouzeniaRepairAgentAutonomousLLMBased2025, leeUnifiedDebuggingApproach2024, yangSWEagentAgentComputerInterfaces2024, zhangAutoCodeRoverAutonomousProgram2024}. Finally, there are hybrid methods that combine various traditional and learning-based strategies, leveraging syntactic patterns, semantic understanding, and neural models to expand the scope of patch generation and verification \citep{mengTemplatebasedNeuralProgram2023, parasaramReteLearningNamespace2023, xiaLessTrainingMore2022, yiSpeedingConstraintbasedProgram2022}. Our work uses a pre-trained code language model that is fine-tuned using supervised learning to generate fixes.

In the following subsections, we provide an overview of recent related work in learning-based APR and various input augmentation techniques.

\subsection{Learning-based APR}

Early approaches adapted neural machine translation (NMT) techniques to transform buggy code into fixes.
\citet{tufanoEmpiricalStudyLearning2019} train an encoder-decoder model for automated Java bug fixing. They mine GitHub for bug-fixing commits, extract method-level buggy-fixed code pairs, and abstract the code to reduce vocabulary size.
SequenceR \citep{chenSequenceRSequencetoSequenceLearning2019} is trained on curated samples from GitHub repositories and evaluated on CodRep \citep{chenCodRepMachineLearning2018} and Defects4J \citep{justDefects4JDatabaseExisting2014}. SequenceR incorporates a copy mechanism to better handle the large vocabulary of source code and an abstract buggy context to improve adaptability to diverse bug types.
CoCoNuT \citep{lutellierCoCoNuTCombiningContextaware2020} uses a context-aware NMT architecture trained on bug-fix pairs from open-source repositories to fix bugs across multiple programming languages, including Java, Python, C, and JavaScript. Its NMT architecture uses separate encoders for the buggy code and its surrounding context, along with a shared decoder to generate fixes. CoCoNuT uses a stack of fully convolutional layers (FConv) in its encoders to effectively extract hierarchical features and model the source code at varying granularity levels. Additionally, CoCoNuT leverages ensemble learning by combining multiple models with different hyperparameters to address different bugs.
\citet{yuanCIRCLEContinualRepair2022} propose CIRCLE that uses continual learning for bug-fixing across multiple programming languages. It employs a prompting function to bridge the gap between NLP pre-trained tasks and APR, a difficulty-based example replay for lifelong learning without requiring full historical data, and a sampling-based elastic weight regularization method to prevent catastrophic forgetting. The approach also includes a re-repairing mechanism to correct cross-language generation errors.

As NMT-based methods matured, studies began incorporating syntax and semantic constraints to improve patch validity.
\citet{jiangCURECodeAwareNeural2021} introduce an APR technique called CURE with three main innovations: pre-training a programming language model based on the GPT architecture \citep{radfordImprovingLanguageUnderstanding2018} on a large codebase, designing a code-aware beam search strategy that focuses on compilable and length-consistent patches, and using subword tokenization to reduce the fix search space.
Recoder \citep{zhuSyntaxguidedEditDecoder2021} presents a syntax-guided edit decoder with placeholder generation to generate syntactically correct patches and handle project-specific identifiers. Instead of a complete sequence of tokens, Recoder generates edits to allow for efficient representation of small changes.
RewardRepair \citep{yeNeuralProgramRepair2022} uses a discriminative model to modulate the cross-entropy loss of the model before backpropagation, rewarding patches that compile and pass test cases while penalizing those that are identical to the buggy code or introduce regression errors.

KNOD \citep{jiangKNODDomainKnowledge2023} incorporates domain knowledge of source code to improve patch generation syntactically and semantically. It features a three-stage tree decoder that generates abstract syntax trees (ASTs) of patched code, preserving the inherent tree structure of source code. Furthermore, KNOD employs domain-rule distillation to enforce syntactic and semantic rules during both the training and inference phases.
\citet{zhuTareTypeAwareNeural2023} propose Tare, a type-aware model designed to learn the typing rules of programming languages and generate more typeable patches. Tare incorporates a specialized grammar called T-Grammar, which integrates type information into standard grammar, along with a novel code representation called T-Graph that captures key information for type-checking ASTs.
\citet{weiCopilotingCopilotsFusing2023} propose Repilot, a code generation framework that integrates a completion engine commonly used in development environments (specifically the Eclipse JDT Language Server) with LLMs to guide them to synthesize more valid patches during the repair process. By synthesizing candidate patches through a collaborative process, Repilot uses the LLM to generate initial token probabilities autoregressively and then queries the completion engine to dynamically prune invalid tokens and complete the partially generated tokens based on its suggestions.

RAP-Gen \citep{wangRAPGenRetrievalAugmentedPatch2023} integrates retrieval-augmented techniques with the CodeT5 language model for patch generation. The authors employ a hybrid sparse and dense patch retriever to extract relevant guiding fix patterns from a codebase of previous bug-fix pairs, leveraging both lexical and semantic matching. These retrieved fix patterns augment the buggy input provided to the CodeT5 patch generator to synthesize a list of repair candidates.
RAP-Gen's approach is more sophisticated than ours with multiple instances of CodeT5 fine-tuned for patch generation and retrieval on code. In contrast, we use a smaller model pre-trained on natural language for retrieval and only use a semantic-based retriever rather than a hybrid one. Future work could explore expanding these retrieval techniques. We also do not use a codebase of buggy-fixed pairs to retrieve examples of how to fix a bug. Instead, we rely on the content available within the buggy source files.

To address complex multi-location bugs, DEAR \citep{liDEARNovelDeep2022} uses a tree-based LSTM model with cycle training and a divide-and-conquer strategy to learn appropriate code transformations. It selects and combines multiple buggy locations to generate patches for all identified locations simultaneously to repair multi-hunk bugs.
\citet{yeITERIterativeNeural2024} propose an iterative paradigm called ITER that focuses on both single-location and multi-location patches. ITER is designed to improve partial patches until they become plausible and correct. To improve the partial patches, ITER repeatedly invokes the model with partial patches in a multistep manner to let the model repair its mistakes.
The second part of our multi-hunk patch generation procedure is inspired by ITER where we similarly combine single-hunk partial patches to form a full patch for a multi-hunk bug. However, our approach follows a simpler one-step patch generation process, where we call the model once for each hunk instead of iteratively invoking it multiple times to refine the generated patches. We also do not necessarily change all hunks and in some cases leave certain hunks unchanged when none of the generated patches for those hunks improve the situation, hoping to find a patch by modifying other hunks.

T5APR \citep{gharibiT5APREmpoweringAutomated2024} formulates multilingual bug fixing using multitask learning, which involves simultaneous fine-tuning of a pre-trained sequence-to-sequence transformer model. T5APR employs a checkpoint ensemble strategy as an efficient way to improve its patch recommendation effectiveness. It also targets multi-hunk bugs that share the same fix across all hunks.
We build \approach{} on top of T5APR with some core modifications to use its codebase. Specifically, while T5APR does not consider any surrounding context when the buggy code is outside a function, \approach{} uses preceding and following lines as context. Furthermore, we extend it with our contributions by context augmentation and multi-hunk fix capabilities.

Several proposed tools combine the power of LLMs with repair templates for APR.
\citet{xiaLessTrainingMore2022} introduce AlphaRepair, a zero-shot approach that uses cloze-style code infilling to generate fixes without requiring large bug-fixing datasets. Instead of predicting explicit repair edits, it leverages the masked language modeling objective of pre-trained models like CodeBERT \citep{fengCodeBERTPreTrainedModel2020}, inferring missing or incorrect code through contextual information. To enhance performance, AlphaRepair employs multiple masking templates on target code lines.
TENURE \citep{mengTemplatebasedNeuralProgram2023} first builds a large-scale dataset of repair templates and then trains encoder-decoder models to learn to generate patches for different templates. It also uses the copy mechanism, similar to SequenceR, to address the out-of-vocabulary problem by replacing unknown tokens with project-specific information.
Gamma \citep{zhangGammaRevisitingTemplateBased2023} collects and revises a variety of fix templates from state-of-the-art template-based APR techniques (e.g., TBar \citep{liuTBarRevisitingTemplatebased2019}) and transforms them into masked hole-filling-based patterns. The authors then adopt a pre-trained language model UniXcoder \citep{guoUniXcoderUnifiedCrossModal2022} to predict the correct code for masked sections as a fill-in-the-blank task, utilizing the surrounding buggy context and the generic knowledge of the model pre-trained with code snippets from open-source projects.

Recent studies investigate the performance of different LLMs for APR, comparing them with each other to assess their relative strengths and weaknesses \citep{huangEmpiricalStudyFineTuning2023}. These studies also consider the conversational capabilities of instruction-tuned models with multiple rounds of dialogue with different prompts and additional information for incorrect patches to improve their performance.
\citet{prennerCanOpenAICodex2022} demonstrates that OpenAI's Codex, a GPT-3-based model trained on a large code corpus, performs competitively with state-of-the-art APR tools on the QuixBugs benchmark, particularly excelling in Python over Java, with prompt design significantly influencing repair success. Similarly, \citet{sobaniaAnalysisAutomaticBug2023} find that ChatGPT rivals Codex and traditional APR approaches. Additionally, providing clarifying hints, such as expected outputs or error messages in its dialogue-based system, further boosts the effectiveness.
\citet{caoStudyPromptDesign2025} explore the impact of prompt design on the debugging performance of ChatGPT and proposes various prompt templates to enhance its performance.

Expanding the scope, \citet{xiaAutomatedProgramRepair2023} perform a systematic study on directly applying nine pre-trained generative and infilling LLMs under several configurations for APR across three repair settings: generating entire functions, filling code chunks based on contextual information, and producing single-line fixes. The results demonstrate that LLMs outperform traditional APR tools and exhibit a scaling effect, where larger models achieve superior performance.
These findings align with those of \citet{jiangImpactCodeLanguage2023}. Complementing these results, they also observe that while unmodified code language models struggle to leverage buggy lines effectively, fine-tuned models tend to over-rely on them.
Finally, \citet{zhangCriticalReviewLarge2024} evaluate ChatGPT on a new APR benchmark from competitive programming problems post-ChatGPT's training cutoff point to prevent benchmark leakage and compare its performance with other LLMs. The study further highlights the potential of interactive conversation-based repair workflows.

\begin{table}
  \caption{Comparison of different learning-based APR approaches.}
  \centering
  \scriptsize
  \tabcolsep=0.2cm
  \begin{tabular}{lccll}
    \toprule
    Approach                                                      & Evaluated language          & Beam size & Tokenizer               & Model                                        \\
    \midrule
    \citet{tufanoEmpiricalStudyLearning2019}                      & Java                        & 50        & Word                    & Encoder-decoder LSTM                         \\
    SequenceR \citep{chenSequenceRSequencetoSequenceLearning2019} & Java                        & 50        & Word                    & Encoder-decoder LSTM                         \\
    DLFix \citep{liDLFixContextbasedCode2020}                     & Java                        & -         & Word                    & Tree-based LSTM                              \\
    CoCoNuT \citep{lutellierCoCoNuTCombiningContextaware2020}     & Java, Python, C, JavaScript & 1,000     & Word                    & FConv                                        \\
    CURE \citep{jiangCURECodeAwareNeural2021}                     & Java                        & 1,000     & Subword (BPE)           & GPT + FConv                                  \\
    Recoder \citep{zhuSyntaxguidedEditDecoder2021}                & Java                        & 100       & Word                    & Tree-based transformer                       \\
    RewardRepair \citep{yeNeuralProgramRepair2022}                & Java                        & 200       & Subword (SentencePiece) & T5-base (220M)                               \\
    CIRCLE \citep{yuanCIRCLEContinualRepair2022}                  & Java, Python, C, JavaScript & 250       & Subword (SentencePiece) & T5-base (220M)                               \\
    \citet{prennerCanOpenAICodex2022}                             & Java, Python                & -         & Subword (BPE)           & Davinci-codex                                \\
    \citet{sobaniaAnalysisAutomaticBug2023}                       & Java                        & -         & Subword (BPE)           & GPT-3.5-turbo                                \\
    KNOD \citep{jiangKNODDomainKnowledge2023}                     & Java                        & 1,000     & Word                    & Graph-based transformer + Tree-based decoder \\
    AlphaRepair \citep{xiaLessTrainingMore2022}                   & Java                        & 25        & Subword (BPE)           & CodeBERT-base                                \\
    SelfAPR \citep{yeSelfAPRSelfsupervisedProgram2023}            & Java                        & 50        & Subword (SentencePiece) & T5-base (220M)                               \\
    RAP-Gen \citep{wangRAPGenRetrievalAugmentedPatch2023}         & Java, JavaScript            & 100       & Subword (BPE)           & CodeT5-base (220M)                           \\
    Repilot \citep{weiCopilotingCopilotsFusing2023}               & Java                        & 5,000     & Subword (BPE)           & CodeT5-large (770M)                          \\
    Tare \citep{zhuTareTypeAwareNeural2023}                       & Java                        & 100       & Word                    & Graph-based transformer                      \\
    TENURE \citep{mengTemplatebasedNeuralProgram2023}             & Java                        & 500       & Word                    & Encoder-decoder LSTM                         \\
    Gamma \citep{zhangGammaRevisitingTemplateBased2023}           & Java                        & 250       & Subword (BPE)           & UniXcoder-base                               \\
    T5APR \citep{gharibiT5APREmpoweringAutomated2024}             & Java, Python, C, JavaScript & 100       & Subword (BPE)           & CodeT5-small (60M)                           \\
    \midrule
    \approach{}                                                   & Java, Python, C, JavaScript & 100       & Subword (BPE)           & CodeT5-small (60M)                           \\
    \bottomrule
  \end{tabular}
  \label{tab:related-work}
\end{table}

\Cref{tab:related-work} summarizes the characteristics of several learning-based APR approaches reviewed here, including those compared with our tool.

\subsection{Input augmentation}

Input augmentation or prompt augmentation for APR, commonly used with instruction-tuned LLMs, refers to techniques that augment the input beyond the original buggy context. These methods aim to increase the likelihood of identifying the correct fix by providing additional contextual information or structured guidance.

Several approaches leverage static program analysis and program slicing to identify semantically relevant repair ingredients.
DeepDebug \citep{drainDeepDebugFixingPython2021} expands its context window beyond the buggy function by adding code skeletons consisting of the function's parent class, imports, signatures, and docstrings.
\citet{zhangNeuralProgramRepair2023} use program dependence analysis by performing program slicing to extract contextual information that has a semantic relationship with the given buggy statement.
\citet{nashidEmbeddingContextCode2023} introduce a flow-augmented graph representation that encodes information about the buggy code and its repair ingredients. This approach uses backward slicing and embeds control and data flow information to capture relevant semantic relationships between program elements.
\citet{sintahaKatanaDualSlicing2023} present dual-slicing, a program slicing approach that extracts bug-relevant code segments by analyzing control and data dependencies in both buggy and fixed code versions, capturing repair context for training.

Augmenting the input with runtime and static analyzer signals is another method used in the literature.
SelfAPR \citep{yeSelfAPRSelfsupervisedProgram2023} executes buggy training samples and extracts and encodes compiler and test execution diagnostics into the input representation of the neural model.
\citet{zhangCriticalReviewLarge2024} also explore the utility of adding problem description, error feedback, and bug localization information into the input prompt.
TFix \citep{berabiTFixLearningFix2021}, a tool that synthesizes fixes for ESLint static analyzer errors in JavaScript, enhances its input context by including the static analyzer's error type and error message.
InferFix \citep{jinInferFixEndEndProgram2023} is another repair tool that addresses static analyzer violations raised by Infer. Inferfix augments its input by adding bug-type annotations of the static checker (e.g., null pointer dereference, resource leak, and thread safety violation), surrounding method with syntactic hierarchies of the focal method, and uses dense retrieval to search for semantically similar fixes from an external database of historical bugs and fixes, blending static analysis with retrieval-augmented generation.

Other tools use retrieval techniques for repair as well.
\citet{nashidRetrievalBasedPromptSelection2023} explore few-shot learning for code-related tasks where they examine lexical and semantic-based retrieval techniques to select code examples from a demonstration pool.
RAP-Gen \citep{wangRAPGenRetrievalAugmentedPatch2023} proposes a hybrid approach combining lexical and semantic techniques to retrieve repair examples.

Some studies systematically experiment with selecting various levels of contextual information.
\citet{prennerOutContextHow2024} investigate the role of local context in the success of neural program repair models. The authors evaluate how much local context (code around the buggy location) is needed for effective bug fixes and whether the context's position impacts repair success. Their study finds that repair success generally increases with larger local context windows, but it is not for all bug types. The analysis of bug patterns reveals that bug types involving identifiers particularly benefit from increased context. They also observe that different context sizes and configurations yield different results, which can be leveraged through ensembling techniques to earn further improvements.

\citet{parasaramFactSelectionProblem2025} frame optimal prompt design as a fact selection problem and explore the construction of effective prompts for LLM-based APR by choosing relevant contextual information from the buggy program and external sources (e.g., stack traces, issue reports, and angelic values) to present to the model during the repair process. Considering the ever-increasing context window of these models, they conduct a large-scale study featuring various combinations of seven diverse facts. Their findings reveal that each fact contributes to fixing some bugs that would otherwise remain unresolved or have a low success rate. Regarding the number of facts used, they discover that while adding facts generally improves performance, there is a threshold beyond which additional facts hinder performance and there is no universal set of facts for all bug repair scenarios. To address this, they propose a statistical model named Maniple to optimize fact selection specific to a given bug.

Our work uniquely investigates line similarity retrieval from buggy source files as a lightweight, structure-agnostic signal for learning-based repair. Although tested on a non-instruction-tuned small model with limited input size, our approach is compatible with instruction-tuned LLMs and complementary to prior work, offering a novel dimension for enriching repair prompts.

\section{Conclusion}
\label{sec:conclusion}
In this paper, we proposed \approach{}, a learning-based approach for multilingual automated program repair (APR).
Our approach leverages retrieval-augmented generation to provide language-independent repair ingredients and fix examples directly from the buggy project files during patch generation, thereby eliminating the dependency on external bug-fixing databases.
We also introduced a systematic methodology to reduce the search space for multi-hunk patch generation, a class of bugs that require fixes across multiple non-contiguous locations in the code.

We conducted a comprehensive evaluation of \approach{} on six benchmarks across four programming languages. Out of 5,501 bugs from all the benchmarks, \approach{} correctly fixed 2,227, with 1,545 matching the ground-truth developer patches. Additionally, it successfully repaired 121 multi-hunk bugs, showcasing its ability to handle complex bugs.
The contributions of \approach{} highlight its potential to enhance software debugging workflows and reduce manual effort.

There are several directions for future work.
First, using models with larger input sizes or instruction-tuned models for conversation-based APR to combine our context augmentation strategy with other available relevant contextual information could enhance context understanding during patch generation. It is also possible to allow the model to adaptively select its required context for each bug from a collection of options through tool use or chain-of-thought prompting \citep{parasaramFactSelectionProblem2025}.
Second, the proposed multi-hunk strategy could be combined with grouping several closely related hunks into a single input \citep{huangEmpiricalStudyFineTuning2023,silvaRepairLLaMAEfficientRepresentations2025}. Analyzing which test cases trigger specific hunks could help cluster related hunks more effectively, improving patch generation and reducing unnecessary computations.
Third, improving the ranking of generated patches remains a crucial challenge. A more effective ranking mechanism could significantly improve the tool's performance by prioritizing correct patches earlier in the validation process.
Finally, exploring the integration of \approach{} into continuous integration pipelines would facilitate its practical adoption, bridging the gap between research and real-world applications. This integration could enable developers to automatically fix bugs as part of their regular workflows.

\section*{Declarations}

\subsection*{Data availability}
The data, code, and trained models used in this study are available at the paper's repository: \url{https://github.com/h4iku/MultiMend}.

\subsection*{Conflict of interest}
The authors declare that they have no conflict of interest.

\bibliographystyle{spbasic}
\bibliography{references}  

\begin{thebibliography}{111}
\providecommand{\natexlab}[1]{#1}
\providecommand{\url}[1]{{#1}}
\providecommand{\urlprefix}{URL }
\expandafter\ifx\csname urlstyle\endcsname\relax
  \providecommand{\doi}[1]{DOI~\discretionary{}{}{}#1}\else
  \providecommand{\doi}{DOI~\discretionary{}{}{}\begingroup \urlstyle{rm}\Url}\fi
\providecommand{\eprint}[2][]{\url{#2}}

\bibitem[{Allamanis et~al(2021)Allamanis, {Jackson-Flux}, and Brockschmidt}]{allamanisSelfSupervisedBugDetection2021}
Allamanis M, {Jackson-Flux} H, Brockschmidt M (2021) Self-{{Supervised Bug Detection}} and {{Repair}}. In: Advances in {{Neural Information Processing Systems}}, Curran Associates, Inc., vol~34, pp 27,865--27,876, \url{https://proceedings.neurips.cc/paper_files/paper/2021/hash/ea96efc03b9a050d895110db8c4af057-Abstract.html}

\bibitem[{An et~al(2023)An, Kwon, Choi, Yi, and Yoo}]{anBugsCHighlyUsable2023}
An G, Kwon M, Choi K, Yi J, Yoo S (2023) {{BugsC}}++: {{A Highly Usable Real World Defect Benchmark}} for {{C}}/{{C}}++. In: 2023 38th {{IEEE}}/{{ACM International Conference}} on {{Automated Software Engineering}} ({{ASE}}), pp 2034--2037, \doi{10.1109/ASE56229.2023.00208}

\bibitem[{Barr et~al(2014)Barr, Brun, Devanbu, Harman, and Sarro}]{barrPlasticSurgeryHypothesis2014}
Barr ET, Brun Y, Devanbu P, Harman M, Sarro F (2014) The plastic surgery hypothesis. In: Proceedings of the 22nd {{ACM SIGSOFT International Symposium}} on {{Foundations}} of {{Software Engineering}}, Association for Computing Machinery, New York, NY, USA, {{FSE}} 2014, pp 306--317, \doi{10.1145/2635868.2635898}

\bibitem[{Berabi et~al(2021)Berabi, He, Raychev, and Vechev}]{berabiTFixLearningFix2021}
Berabi B, He J, Raychev V, Vechev M (2021) {{TFix}}: {{Learning}} to {{Fix Coding Errors}} with a {{Text-to-Text Transformer}}. In: Proceedings of the 38th {{International Conference}} on {{Machine Learning}}, PMLR, pp 780--791, \url{https://proceedings.mlr.press/v139/berabi21a.html}

\bibitem[{Bouzenia et~al(2025)Bouzenia, Devanbu, and Pradel}]{bouzeniaRepairAgentAutonomousLLMBased2025}
Bouzenia I, Devanbu P, Pradel M (2025) {{RepairAgent}}: {{An Autonomous}}, {{LLM-Based Agent}} for {{Program Repair}}. In: 2025 {{IEEE}}/{{ACM}} 47th {{International Conference}} on {{Software Engineering}} ({{ICSE}}), pp 2188--2200, \doi{10.1109/ICSE55347.2025.00157}

\bibitem[{Cao et~al(2025)Cao, Li, Wen, and Cheung}]{caoStudyPromptDesign2025}
Cao J, Li M, Wen M, Cheung SC (2025) A study on prompt design, advantages and limitations of {{ChatGPT}} for deep learning program repair. Automated Software Engineering 32(1):30, \doi{10.1007/s10515-025-00492-x}

\bibitem[{Chen et~al(2017)Chen, Lundberg, and Lee}]{chenCheckpointEnsemblesEnsemble2017}
Chen H, Lundberg S, Lee SI (2017) Checkpoint {{Ensembles}}: {{Ensemble Methods}} from a {{Single Training Process}}. \doi{10.48550/arXiv.1710.03282}, \eprint{1710.03282}

\bibitem[{Chen and Monperrus(2018)}]{chenCodRepMachineLearning2018}
Chen Z, Monperrus M (2018) The {{CodRep Machine Learning}} on {{Source Code Competition}}. \doi{10.48550/arXiv.1807.03200}, \eprint{1807.03200}

\bibitem[{Chen et~al(2019)Chen, Kommrusch, Tufano, Pouchet, Poshyvanyk, and Monperrus}]{chenSequenceRSequencetoSequenceLearning2019}
Chen Z, Kommrusch S, Tufano M, Pouchet LN, Poshyvanyk D, Monperrus M (2019) {{SequenceR}}: {{Sequence-to-Sequence Learning}} for {{End-to-End Program Repair}}. IEEE Transactions on Software Engineering 47(9):1943--1959, \doi{10.1109/TSE.2019.2940179}

\bibitem[{Cheshkov et~al(2024)Cheshkov, Zadorozhny, Levichev, Maslov, and Jaldin}]{cheshkovExploringPotentialConversational2024}
Cheshkov A, Zadorozhny P, Levichev R, Maslov E, Jaldin RF (2024) Exploring the {{Potential}} of {{Conversational Test Suite Based Program Repair}} on {{SWE-bench}}. \doi{10.48550/arXiv.2410.04485}, \eprint{2410.04485}

\bibitem[{Ding et~al(2021)Ding, Ray, Devanbu, and Hellendoorn}]{dingPatchingTranslationData2021}
Ding Y, Ray B, Devanbu P, Hellendoorn VJ (2021) Patching as translation: The data and the metaphor. In: Proceedings of the 35th {{IEEE}}/{{ACM International Conference}} on {{Automated Software Engineering}}, Association for Computing Machinery, New York, NY, USA, {{ASE}} '20, pp 275--286, \doi{10.1145/3324884.3416587}

\bibitem[{Drain et~al(2021)Drain, Clement, Serrato, and Sundaresan}]{drainDeepDebugFixingPython2021}
Drain D, Clement CB, Serrato G, Sundaresan N (2021) {{DeepDebug}}: {{Fixing Python Bugs Using Stack Traces}}, {{Backtranslation}}, and {{Code Skeletons}}. \doi{10.48550/arXiv.2105.09352}, \eprint{2105.09352}

\bibitem[{Durieux et~al(2019)Durieux, Madeiral, Martinez, and Abreu}]{durieuxEmpiricalReviewJava2019}
Durieux T, Madeiral F, Martinez M, Abreu R (2019) Empirical review of {{Java}} program repair tools: A large-scale experiment on 2,141 bugs and 23,551 repair attempts. In: Proceedings of the 2019 27th {{ACM Joint Meeting}} on {{European Software Engineering Conference}} and {{Symposium}} on the {{Foundations}} of {{Software Engineering}}, Association for Computing Machinery, New York, NY, USA, {{ESEC}}/{{FSE}} 2019, pp 302--313, \doi{10.1145/3338906.3338911}

\bibitem[{Feng et~al(2020)Feng, Guo, Tang, Duan, Feng, Gong, Shou, Qin, Liu, Jiang, and Zhou}]{fengCodeBERTPreTrainedModel2020}
Feng Z, Guo D, Tang D, Duan N, Feng X, Gong M, Shou L, Qin B, Liu T, Jiang D, Zhou M (2020) {{CodeBERT}}: {{A Pre-Trained Model}} for {{Programming}} and {{Natural Languages}}. In: Findings of the {{Association}} for {{Computational Linguistics}}: {{EMNLP}} 2020, Association for Computational Linguistics, Online, pp 1536--1547, \doi{10.18653/v1/2020.findings-emnlp.139}

\bibitem[{Gao et~al(2022)Gao, Noller, and Roychoudhury}]{gaoProgramRepair2022}
Gao X, Noller Y, Roychoudhury A (2022) Program {{Repair}}. \doi{10.48550/arXiv.2211.12787}, \eprint{2211.12787}

\bibitem[{Gao et~al(2024)Gao, Xiong, Gao, Jia, Pan, Bi, Dai, Sun, Wang, and Wang}]{gaoRetrievalAugmentedGenerationLarge2024}
Gao Y, Xiong Y, Gao X, Jia K, Pan J, Bi Y, Dai Y, Sun J, Wang M, Wang H (2024) Retrieval-{{Augmented Generation}} for {{Large Language Models}}: {{A Survey}}. \doi{10.48550/arXiv.2312.10997}, \eprint{2312.10997}

\bibitem[{Gharibi et~al(2018)Gharibi, Rasekh, Sadreddini, and Fakhrahmad}]{gharibiLeveragingTextualProperties2018}
Gharibi R, Rasekh AH, Sadreddini MH, Fakhrahmad SM (2018) Leveraging textual properties of bug reports to localize relevant source files. Information Processing \& Management 54(6):1058--1076, \doi{10.1016/j.ipm.2018.07.004}

\bibitem[{Gharibi et~al(2024)Gharibi, Sadreddini, and Fakhrahmad}]{gharibiT5APREmpoweringAutomated2024}
Gharibi R, Sadreddini MH, Fakhrahmad SM (2024) {{T5APR}}: {{Empowering}} automated program repair across languages through checkpoint ensemble. Journal of Systems and Software 214:112,083, \doi{10.1016/j.jss.2024.112083}

\bibitem[{Guo et~al(2022)Guo, Lu, Duan, Wang, Zhou, and Yin}]{guoUniXcoderUnifiedCrossModal2022}
Guo D, Lu S, Duan N, Wang Y, Zhou M, Yin J (2022) {{UniXcoder}}: {{Unified Cross-Modal Pre-training}} for {{Code Representation}}. In: Muresan S, Nakov P, Villavicencio A (eds) Proceedings of the 60th {{Annual Meeting}} of the {{Association}} for {{Computational Linguistics}} ({{Volume}} 1: {{Long Papers}}), Association for Computational Linguistics, Dublin, Ireland, pp 7212--7225, \doi{10.18653/v1/2022.acl-long.499}

\bibitem[{Gyimesi et~al(2019)Gyimesi, Vancsics, Stocco, Mazinanian, Besz{\'e}des, Ferenc, and Mesbah}]{gyimesiBugsJSBenchmarkJavaScript2019}
Gyimesi P, Vancsics B, Stocco A, Mazinanian D, Besz{\'e}des {\'A}, Ferenc R, Mesbah A (2019) {{BugsJS}}: A {{Benchmark}} of {{JavaScript Bugs}}. In: 2019 12th {{IEEE Conference}} on {{Software Testing}}, {{Validation}} and {{Verification}} ({{ICST}}), pp 90--101, \doi{10.1109/ICST.2019.00019}

\bibitem[{Hamill and {Goseva-Popstojanova}(2017)}]{hamillAnalyzingPredictingEffort2017}
Hamill M, {Goseva-Popstojanova} K (2017) Analyzing and predicting effort associated with finding and fixing software faults. Information and Software Technology 87:1--18, \doi{10.1016/j.infsof.2017.01.002}

\bibitem[{Hanam et~al(2016)Hanam, Brito, and Mesbah}]{hanamDiscoveringBugPatterns2016}
Hanam Q, Brito FSdM, Mesbah A (2016) Discovering bug patterns in {{JavaScript}}. In: Proceedings of the 2016 24th {{ACM SIGSOFT International Symposium}} on {{Foundations}} of {{Software Engineering}}, Association for Computing Machinery, New York, NY, USA, {{FSE}} 2016, pp 144--156, \doi{10.1145/2950290.2950308}

\bibitem[{Huang et~al(2017)Huang, Li, Pleiss, Liu, Hopcroft, and Weinberger}]{huangSnapshotEnsemblesTrain2017}
Huang G, Li Y, Pleiss G, Liu Z, Hopcroft JE, Weinberger KQ (2017) Snapshot {{Ensembles}}: {{Train}} 1, {{Get M}} for {{Free}}. In: International {{Conference}} on {{Learning Representations}}, \url{https://openreview.net/forum?id=BJYwwY9ll}

\bibitem[{Huang et~al(2023{\natexlab{a}})Huang, Meng, Zhang, Liu, Wang, Li, and Zhang}]{huangEmpiricalStudyFineTuning2023}
Huang K, Meng X, Zhang J, Liu Y, Wang W, Li S, Zhang Y (2023{\natexlab{a}}) An {{Empirical Study}} on {{Fine-Tuning Large Language Models}} of {{Code}} for {{Automated Program Repair}}. In: 2023 38th {{IEEE}}/{{ACM International Conference}} on {{Automated Software Engineering}} ({{ASE}}), pp 1162--1174, \doi{10.1109/ASE56229.2023.00181}

\bibitem[{Huang et~al(2023{\natexlab{b}})Huang, Xu, Yang, Sun, Li, Yan, and Zhang}]{huangSurveyAutomatedProgram2023}
Huang K, Xu Z, Yang S, Sun H, Li X, Yan Z, Zhang Y (2023{\natexlab{b}}) A {{Survey}} on {{Automated Program Repair Techniques}}. \doi{10.48550/arXiv.2303.18184}, \eprint{2303.18184}

\bibitem[{Husain et~al(2020)Husain, Wu, Gazit, Allamanis, and Brockschmidt}]{husainCodeSearchNetChallengeEvaluating2020}
Husain H, Wu HH, Gazit T, Allamanis M, Brockschmidt M (2020) {{CodeSearchNet Challenge}}: {{Evaluating}} the {{State}} of {{Semantic Code Search}}. \doi{10.48550/arXiv.1909.09436}, \eprint{1909.09436}

\bibitem[{Jiang et~al(2018)Jiang, Xiong, Zhang, Gao, and Chen}]{jiangShapingProgramRepair2018}
Jiang J, Xiong Y, Zhang H, Gao Q, Chen X (2018) Shaping program repair space with existing patches and similar code. In: Proceedings of the 27th {{ACM SIGSOFT International Symposium}} on {{Software Testing}} and {{Analysis}}, Association for Computing Machinery, New York, NY, USA, {{ISSTA}} 2018, pp 298--309, \doi{10.1145/3213846.3213871}

\bibitem[{Jiang et~al(2021)Jiang, Lutellier, and Tan}]{jiangCURECodeAwareNeural2021}
Jiang N, Lutellier T, Tan L (2021) {{CURE}}: {{Code-Aware Neural Machine Translation}} for {{Automatic Program Repair}}. In: 2021 {{IEEE}}/{{ACM}} 43rd {{International Conference}} on {{Software Engineering}} ({{ICSE}}), pp 1161--1173, \doi{10.1109/ICSE43902.2021.00107}

\bibitem[{Jiang et~al(2023{\natexlab{a}})Jiang, Liu, Lutellier, and Tan}]{jiangImpactCodeLanguage2023}
Jiang N, Liu K, Lutellier T, Tan L (2023{\natexlab{a}}) Impact of {{Code Language Models}} on {{Automated Program Repair}}. In: 2023 {{IEEE}}/{{ACM}} 45th {{International Conference}} on {{Software Engineering}} ({{ICSE}}), pp 1430--1442, \doi{10.1109/ICSE48619.2023.00125}

\bibitem[{Jiang et~al(2023{\natexlab{b}})Jiang, Lutellier, Lou, Tan, Goldwasser, and Zhang}]{jiangKNODDomainKnowledge2023}
Jiang N, Lutellier T, Lou Y, Tan L, Goldwasser D, Zhang X (2023{\natexlab{b}}) {{KNOD}}: {{Domain Knowledge Distilled Tree Decoder}} for {{Automated Program Repair}}. In: Proceedings of the 45th {{International Conference}} on {{Software Engineering}}, IEEE Press, Melbourne, Victoria, Australia, {{ICSE}} '23, pp 1251--1263, \doi{10.1109/ICSE48619.2023.00111}

\bibitem[{Jin et~al(2023)Jin, Shahriar, Tufano, Shi, Lu, Sundaresan, and Svyatkovskiy}]{jinInferFixEndEndProgram2023}
Jin M, Shahriar S, Tufano M, Shi X, Lu S, Sundaresan N, Svyatkovskiy A (2023) {{InferFix}}: {{End-to-End Program Repair}} with {{LLMs}}. In: Proceedings of the 31st {{ACM Joint European Software Engineering Conference}} and {{Symposium}} on the {{Foundations}} of {{Software Engineering}}, Association for Computing Machinery, New York, NY, USA, {{ESEC}}/{{FSE}} 2023, pp 1646--1656, \doi{10.1145/3611643.3613892}

\bibitem[{Just et~al(2014)Just, Jalali, and Ernst}]{justDefects4JDatabaseExisting2014}
Just R, Jalali D, Ernst MD (2014) {{Defects4J}}: A database of existing faults to enable controlled testing studies for {{Java}} programs. In: Proceedings of the 2014 {{International Symposium}} on {{Software Testing}} and {{Analysis}}, Association for Computing Machinery, New York, NY, USA, {{ISSTA}} 2014, pp 437--440, \doi{10.1145/2610384.2628055}

\bibitem[{Kong et~al(2025)Kong, Xie, Cheng, Liu, Du, and Guo}]{kongContrastRepairEnhancingConversationBased2025}
Kong J, Xie X, Cheng M, Liu S, Du X, Guo Q (2025) {{ContrastRepair}}: {{Enhancing Conversation-Based Automated Program Repair}} via {{Contrastive Test Case Pairs}}. ACM Trans Softw Eng Methodol \doi{10.1145/3719345}

\bibitem[{Koyuncu et~al(2020)Koyuncu, Liu, Bissyand{\'e}, Kim, Klein, Monperrus, and Le~Traon}]{koyuncuFixMinerMiningRelevant2020}
Koyuncu A, Liu K, Bissyand{\'e} TF, Kim D, Klein J, Monperrus M, Le~Traon Y (2020) {{FixMiner}}: {{Mining}} relevant fix patterns for automated program repair. Empirical Software Engineering 25(3):1980--2024, \doi{10.1007/s10664-019-09780-z}

\bibitem[{Le~Goues et~al(2012)Le~Goues, Nguyen, Forrest, and Weimer}]{legouesGenProgGenericMethod2012}
Le~Goues C, Nguyen T, Forrest S, Weimer W (2012) {{GenProg}}: {{A Generic Method}} for {{Automatic Software Repair}}. IEEE Transactions on Software Engineering 38(1):54--72, \doi{10.1109/TSE.2011.104}

\bibitem[{Le~Goues et~al(2019)Le~Goues, Pradel, and Roychoudhury}]{legouesAutomatedProgramRepair2019}
Le~Goues C, Pradel M, Roychoudhury A (2019) Automated program repair. Communications of the ACM 62(12):56--65, \doi{10.1145/3318162}

\bibitem[{Lee et~al(2024)Lee, Xia, Yang, Huang, Zhu, Zhang, and Lyu}]{leeUnifiedDebuggingApproach2024}
Lee C, Xia CS, Yang L, Huang Jt, Zhu Z, Zhang L, Lyu MR (2024) A {{Unified Debugging Approach}} via {{LLM-Based Multi-Agent Synergy}}. \doi{10.48550/arXiv.2404.17153}, \eprint{2404.17153}

\bibitem[{Lewis et~al(2020)Lewis, Perez, Piktus, Petroni, Karpukhin, Goyal, K{\"u}ttler, Lewis, Yih, Rockt{\"a}schel, Riedel, and Kiela}]{lewisRetrievalAugmentedGenerationKnowledgeIntensive2020}
Lewis P, Perez E, Piktus A, Petroni F, Karpukhin V, Goyal N, K{\"u}ttler H, Lewis M, Yih Wt, Rockt{\"a}schel T, Riedel S, Kiela D (2020) Retrieval-{{Augmented Generation}} for {{Knowledge-Intensive NLP Tasks}}. In: Advances in {{Neural Information Processing Systems}}, Curran Associates, Inc., vol~33, pp 9459--9474, \url{https://proceedings.neurips.cc/paper/2020/hash/6b493230205f780e1bc26945df7481e5-Abstract.html}

\bibitem[{Lhoest et~al(2021)Lhoest, {Villanova del Moral}, Jernite, Thakur, {von Platen}, Patil, Chaumond, Drame, Plu, Tunstall, Davison, {\v S}a{\v s}ko, Chhablani, Malik, Brandeis, Le~Scao, Sanh, Xu, Patry, {McMillan-Major}, Schmid, Gugger, Delangue, Matussi{\`e}re, Debut, Bekman, Cistac, Goehringer, Mustar, Lagunas, Rush, and Wolf}]{lhoestDatasetsCommunityLibrary2021}
Lhoest Q, {Villanova del Moral} A, Jernite Y, Thakur A, {von Platen} P, Patil S, Chaumond J, Drame M, Plu J, Tunstall L, Davison J, {\v S}a{\v s}ko M, Chhablani G, Malik B, Brandeis S, Le~Scao T, Sanh V, Xu C, Patry N, {McMillan-Major} A, Schmid P, Gugger S, Delangue C, Matussi{\`e}re T, Debut L, Bekman S, Cistac P, Goehringer T, Mustar V, Lagunas F, Rush A, Wolf T (2021) Datasets: {{A Community Library}} for {{Natural Language Processing}}. In: Proceedings of the 2021 {{Conference}} on {{Empirical Methods}} in {{Natural Language Processing}}: {{System Demonstrations}}, Association for Computational Linguistics, Online and Punta Cana, Dominican Republic, pp 175--184, \doi{10.18653/v1/2021.emnlp-demo.21}

\bibitem[{Li et~al(2020)Li, Wang, and Nguyen}]{liDLFixContextbasedCode2020}
Li Y, Wang S, Nguyen TN (2020) {{DLFix}}: Context-based code transformation learning for automated program repair. In: Proceedings of the {{ACM}}/{{IEEE}} 42nd {{International Conference}} on {{Software Engineering}}, Association for Computing Machinery, New York, NY, USA, {{ICSE}} '20, pp 602--614, \doi{10.1145/3377811.3380345}

\bibitem[{Li et~al(2022)Li, Wang, and Nguyen}]{liDEARNovelDeep2022}
Li Y, Wang S, Nguyen TN (2022) {{DEAR}}: A novel deep learning-based approach for automated program repair. In: Proceedings of the 44th {{International Conference}} on {{Software Engineering}}, Association for Computing Machinery, New York, NY, USA, {{ICSE}} '22, pp 511--523, \doi{10.1145/3510003.3510177}

\bibitem[{Lin et~al(2017)Lin, Koppel, Chen, and {Solar-Lezama}}]{linQuixBugsMultilingualProgram2017}
Lin D, Koppel J, Chen A, {Solar-Lezama} A (2017) {{QuixBugs}}: A multi-lingual program repair benchmark set based on the quixey challenge. In: Proceedings {{Companion}} of the 2017 {{ACM SIGPLAN International Conference}} on {{Systems}}, {{Programming}}, {{Languages}}, and {{Applications}}: {{Software}} for {{Humanity}}, Association for Computing Machinery, New York, NY, USA, {{SPLASH Companion}} 2017, pp 55--56, \doi{10.1145/3135932.3135941}

\bibitem[{Liu et~al(2019{\natexlab{a}})Liu, Koyuncu, Bissyand{\'e}, Kim, Klein, and Le~Traon}]{liuYouCannotFix2019}
Liu K, Koyuncu A, Bissyand{\'e} TF, Kim D, Klein J, Le~Traon Y (2019{\natexlab{a}}) You {{Cannot Fix What You Cannot Find}}! {{An Investigation}} of {{Fault Localization Bias}} in {{Benchmarking Automated Program Repair Systems}}. In: 2019 12th {{IEEE Conference}} on {{Software Testing}}, {{Validation}} and {{Verification}} ({{ICST}}), pp 102--113, \doi{10.1109/ICST.2019.00020}

\bibitem[{Liu et~al(2019{\natexlab{b}})Liu, Koyuncu, Kim, and Bissyand{\'e}}]{liuTBarRevisitingTemplatebased2019}
Liu K, Koyuncu A, Kim D, Bissyand{\'e} TF (2019{\natexlab{b}}) {{TBar}}: Revisiting template-based automated program repair. In: Proceedings of the 28th {{ACM SIGSOFT International Symposium}} on {{Software Testing}} and {{Analysis}}, Association for Computing Machinery, New York, NY, USA, {{ISSTA}} 2019, pp 31--42, \doi{10.1145/3293882.3330577}

\bibitem[{Liu et~al(2020)Liu, Wang, Koyuncu, Kim, Bissyand{\'e}, Kim, Wu, Klein, Mao, and Traon}]{liuEfficiencyTestSuite2020}
Liu K, Wang S, Koyuncu A, Kim K, Bissyand{\'e} TF, Kim D, Wu P, Klein J, Mao X, Traon YL (2020) On the efficiency of test suite based program repair: {{A Systematic Assessment}} of 16 {{Automated Repair Systems}} for {{Java Programs}}. In: Proceedings of the {{ACM}}/{{IEEE}} 42nd {{International Conference}} on {{Software Engineering}}, Association for Computing Machinery, New York, NY, USA, {{ICSE}} '20, pp 615--627, \doi{10.1145/3377811.3380338}

\bibitem[{Loshchilov and Hutter(2018)}]{loshchilovDecoupledWeightDecay2018}
Loshchilov I, Hutter F (2018) Decoupled {{Weight Decay Regularization}}. In: International {{Conference}} on {{Learning Representations}}, \url{https://openreview.net/forum?id=Bkg6RiCqY7}

\bibitem[{Lutellier et~al(2020)Lutellier, Pham, Pang, Li, Wei, and Tan}]{lutellierCoCoNuTCombiningContextaware2020}
Lutellier T, Pham HV, Pang L, Li Y, Wei M, Tan L (2020) {{CoCoNuT}}: Combining context-aware neural translation models using ensemble for program repair. In: Proceedings of the 29th {{ACM SIGSOFT International Symposium}} on {{Software Testing}} and {{Analysis}}, Association for Computing Machinery, New York, NY, USA, {{ISSTA}} 2020, pp 101--114, \doi{10.1145/3395363.3397369}

\bibitem[{Madeiral and Durieux(2021)}]{madeiralLargescaleStudyHumancloned2021}
Madeiral F, Durieux T (2021) A large-scale study on human-cloned changes for automated program repair. In: 2021 {{IEEE}}/{{ACM}} 18th {{International Conference}} on {{Mining Software Repositories}} ({{MSR}}), pp 510--514, \doi{10.1109/MSR52588.2021.00064}

\bibitem[{Mechtaev et~al(2016)Mechtaev, Yi, and Roychoudhury}]{mechtaevAngelixScalableMultiline2016}
Mechtaev S, Yi J, Roychoudhury A (2016) Angelix: Scalable multiline program patch synthesis via symbolic analysis. In: Proceedings of the 38th {{International Conference}} on {{Software Engineering}}, Association for Computing Machinery, New York, NY, USA, {{ICSE}} '16, pp 691--701, \doi{10.1145/2884781.2884807}

\bibitem[{Meng et~al(2023)Meng, Wang, Zhang, Sun, Liu, and Hu}]{mengTemplatebasedNeuralProgram2023}
Meng X, Wang X, Zhang H, Sun H, Liu X, Hu C (2023) Template-based {{Neural Program Repair}}. In: 2023 {{IEEE}}/{{ACM}} 45th {{International Conference}} on {{Software Engineering}} ({{ICSE}}), pp 1456--1468, \doi{10.1109/ICSE48619.2023.00127}

\bibitem[{{Mockus} and {Votta}(2000)}]{mockusIdentifyingReasonsSoftware2000}
{Mockus}, {Votta} (2000) Identifying reasons for software changes using historic databases. In: Proceedings 2000 {{International Conference}} on {{Software Maintenance}}, pp 120--130, \doi{10.1109/ICSM.2000.883028}

\bibitem[{Nashid et~al(2023{\natexlab{a}})Nashid, Sintaha, and Mesbah}]{nashidEmbeddingContextCode2023}
Nashid N, Sintaha M, Mesbah A (2023{\natexlab{a}}) Embedding {{Context}} as {{Code Dependencies}} for {{Neural Program Repair}}. In: 2023 {{IEEE Conference}} on {{Software Testing}}, {{Verification}} and {{Validation}} ({{ICST}}), pp 95--106, \doi{10.1109/ICST57152.2023.00018}

\bibitem[{Nashid et~al(2023{\natexlab{b}})Nashid, Sintaha, and Mesbah}]{nashidRetrievalBasedPromptSelection2023}
Nashid N, Sintaha M, Mesbah A (2023{\natexlab{b}}) Retrieval-{{Based Prompt Selection}} for {{Code-Related Few-Shot Learning}}. In: Proceedings of the 45th {{International Conference}} on {{Software Engineering}}, IEEE Press, Melbourne, Victoria, Australia, {{ICSE}} '23, pp 2450--2462, \doi{10.1109/ICSE48619.2023.00205}

\bibitem[{Nguyen et~al(2013)Nguyen, Qi, Roychoudhury, and Chandra}]{nguyenSemFixProgramRepair2013}
Nguyen HDT, Qi D, Roychoudhury A, Chandra S (2013) {{SemFix}}: {{Program}} repair via semantic analysis. In: 2013 35th {{International Conference}} on {{Software Engineering}} ({{ICSE}}), pp 772--781, \doi{10.1109/ICSE.2013.6606623}

\bibitem[{Nguyen et~al(2019)Nguyen, Ta, and Chin}]{nguyenAutomaticProgramRepair2019}
Nguyen TT, Ta QT, Chin WN (2019) Automatic {{Program Repair Using Formal Verification}} and {{Expression Templates}}. In: Enea C, Piskac R (eds) Verification, {{Model Checking}}, and {{Abstract Interpretation}}, Springer International Publishing, Cham, Lecture {{Notes}} in {{Computer Science}}, pp 70--91, \doi{10.1007/978-3-030-11245-5_4}

\bibitem[{Parasaram et~al(2023)Parasaram, Barr, and Mechtaev}]{parasaramReteLearningNamespace2023}
Parasaram N, Barr ET, Mechtaev S (2023) Rete: {{Learning Namespace Representation}} for {{Program Repair}}. In: 2023 {{IEEE}}/{{ACM}} 45th {{International Conference}} on {{Software Engineering}} ({{ICSE}}), pp 1264--1276, \doi{10.1109/ICSE48619.2023.00112}

\bibitem[{Parasaram et~al(2025)Parasaram, Yan, Yang, Flahy, Qudsi, Ziaber, Barr, and Mechtaev}]{parasaramFactSelectionProblem2025}
Parasaram N, Yan H, Yang B, Flahy Z, Qudsi A, Ziaber D, Barr ET, Mechtaev S (2025) The {{Fact Selection Problem}} in {{LLM-Based Program Repair}}. In: 2025 {{IEEE}}/{{ACM}} 47th {{International Conference}} on {{Software Engineering}} ({{ICSE}}), pp 2574--2586, \doi{10.1109/ICSE55347.2025.00162}

\bibitem[{Paszke et~al(2019)Paszke, Gross, Massa, Lerer, Bradbury, Chanan, Killeen, Lin, Gimelshein, Antiga, Desmaison, Kopf, Yang, DeVito, Raison, Tejani, Chilamkurthy, Steiner, Fang, Bai, and Chintala}]{paszkePyTorchImperativeStyle2019}
Paszke A, Gross S, Massa F, Lerer A, Bradbury J, Chanan G, Killeen T, Lin Z, Gimelshein N, Antiga L, Desmaison A, Kopf A, Yang E, DeVito Z, Raison M, Tejani A, Chilamkurthy S, Steiner B, Fang L, Bai J, Chintala S (2019) {{PyTorch}}: {{An Imperative Style}}, {{High-Performance Deep Learning Library}}. In: Advances in {{Neural Information Processing Systems}}, Curran Associates, Inc., vol~32, \url{https://proceedings.neurips.cc/paper_files/paper/2019/hash/bdbca288fee7f92f2bfa9f7012727740-Abstract.html}

\bibitem[{Prenner and Robbes(2023)}]{prennerRunBugRunExecutableDataset2023}
Prenner JA, Robbes R (2023) {{RunBugRun}} -- {{An Executable Dataset}} for {{Automated Program Repair}}. \doi{10.48550/arXiv.2304.01102}, \eprint{2304.01102}

\bibitem[{Prenner and Robbes(2024)}]{prennerOutContextHow2024}
Prenner JA, Robbes R (2024) Out of {{Context}}: {{How}} important is {{Local Context}} in {{Neural Program Repair}}? In: Proceedings of the {{IEEE}}/{{ACM}} 46th {{International Conference}} on {{Software Engineering}}, Association for Computing Machinery, New York, NY, USA, {{ICSE}} '24, pp 1--13, \doi{10.1145/3597503.3639086}

\bibitem[{Prenner et~al(2022)Prenner, Babii, and Robbes}]{prennerCanOpenAICodex2022}
Prenner JA, Babii H, Robbes R (2022) Can {{OpenAI}}'s codex fix bugs? an evaluation on {{QuixBugs}}. In: Proceedings of the {{Third International Workshop}} on {{Automated Program Repair}}, Association for Computing Machinery, New York, NY, USA, {{APR}} '22, pp 69--75, \doi{10.1145/3524459.3527351}

\bibitem[{Qi et~al(2013)Qi, Mao, Lei, and Wang}]{qiUsingAutomatedProgram2013}
Qi Y, Mao X, Lei Y, Wang C (2013) Using automated program repair for evaluating the effectiveness of fault localization techniques. In: Proceedings of the 2013 {{International Symposium}} on {{Software Testing}} and {{Analysis}}, Association for Computing Machinery, New York, NY, USA, {{ISSTA}} 2013, pp 191--201, \doi{10.1145/2483760.2483785}

\bibitem[{Qi et~al(2015)Qi, Long, Achour, and Rinard}]{qiAnalysisPatchPlausibility2015}
Qi Z, Long F, Achour S, Rinard M (2015) An analysis of patch plausibility and correctness for generate-and-validate patch generation systems. In: Proceedings of the 2015 {{International Symposium}} on {{Software Testing}} and {{Analysis}}, Association for Computing Machinery, New York, NY, USA, {{ISSTA}} 2015, pp 24--36, \doi{10.1145/2771783.2771791}

\bibitem[{Radford et~al(2018)Radford, Narasimhan, Salimans, and Sutskever}]{radfordImprovingLanguageUnderstanding2018}
Radford A, Narasimhan K, Salimans T, Sutskever I (2018) Improving {{Language Understanding}} by {{Generative Pre-Training}}. \url{https://openai.com/index/language-unsupervised/}

\bibitem[{Raffel et~al(2020)Raffel, Shazeer, Roberts, Lee, Narang, Matena, Zhou, Li, and Liu}]{raffelExploringLimitsTransfer2020}
Raffel C, Shazeer N, Roberts A, Lee K, Narang S, Matena M, Zhou Y, Li W, Liu PJ (2020) Exploring the limits of transfer learning with a unified text-to-text transformer. The Journal of Machine Learning Research 21(1):140:5485--140:5551, \url{http://jmlr.org/papers/v21/20-074.html}

\bibitem[{Reimers and Gurevych(2019)}]{reimersSentenceBERTSentenceEmbeddings2019}
Reimers N, Gurevych I (2019) Sentence-{{BERT}}: {{Sentence Embeddings}} using {{Siamese BERT-Networks}}. In: Inui K, Jiang J, Ng V, Wan X (eds) Proceedings of the 2019 {{Conference}} on {{Empirical Methods}} in {{Natural Language Processing}} and the 9th {{International Joint Conference}} on {{Natural Language Processing}} ({{EMNLP-IJCNLP}}), Association for Computational Linguistics, Hong Kong, China, pp 3982--3992, \doi{10.18653/v1/D19-1410}

\bibitem[{Saha et~al(2019)Saha, k.~Saha, and r.~Prasad}]{sahaHarnessingEvolutionMultiHunk2019}
Saha S, k~Saha R, r~Prasad M (2019) Harnessing {{Evolution}} for {{Multi-Hunk Program Repair}}. In: 2019 {{IEEE}}/{{ACM}} 41st {{International Conference}} on {{Software Engineering}} ({{ICSE}}), pp 13--24, \doi{10.1109/ICSE.2019.00020}

\bibitem[{Sennrich et~al(2016)Sennrich, Haddow, and Birch}]{sennrichNeuralMachineTranslation2016}
Sennrich R, Haddow B, Birch A (2016) Neural {{Machine Translation}} of {{Rare Words}} with {{Subword Units}}. In: Proceedings of the 54th {{Annual Meeting}} of the {{Association}} for {{Computational Linguistics}}, Association for Computational Linguistics (ACL), pp 1715--1725, \doi{10.18653/v1/P16-1162}

\bibitem[{Silva et~al(2023)Silva, Ferreira, Ye, and Monperrus}]{silvaMUFINImprovingNeural2023}
Silva A, Ferreira JF, Ye H, Monperrus M (2023) {{MUFIN}}: {{Improving Neural Repair Models}} with {{Back-Translation}}. \doi{10.48550/arXiv.2304.02301}, \eprint{2304.02301}

\bibitem[{Silva et~al(2024)Silva, Saavedra, and Monperrus}]{silvaGitBugJavaReproducibleBenchmark2024}
Silva A, Saavedra N, Monperrus M (2024) {{GitBug-Java}}: {{A Reproducible Benchmark}} of {{Recent Java Bugs}}. In: 2024 {{IEEE}}/{{ACM}} 21st {{International Conference}} on {{Mining Software Repositories}} ({{MSR}}), pp 118--122, \url{https://ieeexplore.ieee.org/abstract/document/10555595}

\bibitem[{Silva et~al(2025)Silva, Fang, and Monperrus}]{silvaRepairLLaMAEfficientRepresentations2025}
Silva A, Fang S, Monperrus M (2025) {{RepairLLaMA}}: {{Efficient Representations}} and {{Fine-Tuned Adapters}} for {{Program Repair}}. IEEE Transactions on Software Engineering 51(8):2366--2380, \doi{10.1109/TSE.2025.3581062}

\bibitem[{Sintaha et~al(2023)Sintaha, Nashid, and Mesbah}]{sintahaKatanaDualSlicing2023}
Sintaha M, Nashid N, Mesbah A (2023) Katana: {{Dual Slicing Based Context}} for {{Learning Bug Fixes}}. ACM Transactions on Software Engineering and Methodology 32(4):100:1--100:27, \doi{10.1145/3579640}

\bibitem[{Smith et~al(2015)Smith, Barr, Le~Goues, and Brun}]{smithCureWorseDisease2015}
Smith EK, Barr ET, Le~Goues C, Brun Y (2015) Is the cure worse than the disease? overfitting in automated program repair. In: Proceedings of the 2015 10th {{Joint Meeting}} on {{Foundations}} of {{Software Engineering}}, Association for Computing Machinery, New York, NY, USA, {{ESEC}}/{{FSE}} 2015, pp 532--543, \doi{10.1145/2786805.2786825}

\bibitem[{Sobania et~al(2023)Sobania, Briesch, Hanna, and Petke}]{sobaniaAnalysisAutomaticBug2023}
Sobania D, Briesch M, Hanna C, Petke J (2023) An {{Analysis}} of the {{Automatic Bug Fixing Performance}} of {{ChatGPT}}. In: 2023 {{IEEE}}/{{ACM International Workshop}} on {{Automated Program Repair}} ({{APR}}), pp 23--30, \doi{10.1109/APR59189.2023.00012}

\bibitem[{Sobreira et~al(2018)Sobreira, Durieux, Madeiral, Monperrus, and {de Almeida Maia}}]{sobreiraDissectionBugDataset2018}
Sobreira V, Durieux T, Madeiral F, Monperrus M, {de Almeida Maia} M (2018) Dissection of a bug dataset: {{Anatomy}} of 395 patches from {{Defects4J}}. In: 2018 {{IEEE}} 25th {{International Conference}} on {{Software Analysis}}, {{Evolution}} and {{Reengineering}} ({{SANER}}), pp 130--140, \doi{10.1109/SANER.2018.8330203}

\bibitem[{Tan et~al(2017)Tan, Yi, {Yulis}, Mechtaev, and Roychoudhury}]{tanCodeflawsProgrammingCompetition2017}
Tan SH, Yi J, {Yulis}, Mechtaev S, Roychoudhury A (2017) Codeflaws: A programming competition benchmark for evaluating automated program repair tools. In: 2017 {{IEEE}}/{{ACM}} 39th {{International Conference}} on {{Software Engineering Companion}} ({{ICSE-C}}), pp 180--182, \doi{10.1109/ICSE-C.2017.76}

\bibitem[{Tian et~al(2023)Tian, Liu, Li, Kabor{\'e}, Koyuncu, Habib, Li, Wen, Klein, and Bissyand{\'e}}]{tianBestBothWorlds2023}
Tian H, Liu K, Li Y, Kabor{\'e} AK, Koyuncu A, Habib A, Li L, Wen J, Klein J, Bissyand{\'e} TF (2023) The {{Best}} of {{Both Worlds}}: {{Combining Learned Embeddings}} with {{Engineered Features}} for {{Accurate Prediction}} of {{Correct Patches}}. ACM Transactions on Software Engineering and Methodology 32(4):92:1--92:34, \doi{10.1145/3576039}

\bibitem[{Tufano et~al(2019{\natexlab{a}})Tufano, Pantiuchina, Watson, Bavota, and Poshyvanyk}]{tufanoLearningMeaningfulCode2019}
Tufano M, Pantiuchina J, Watson C, Bavota G, Poshyvanyk D (2019{\natexlab{a}}) On learning meaningful code changes via neural machine translation. In: Proceedings of the 41st {{International Conference}} on {{Software Engineering}}, IEEE Press, Montreal, Quebec, Canada, {{ICSE}} '19, pp 25--36, \doi{10.1109/ICSE.2019.00021}

\bibitem[{Tufano et~al(2019{\natexlab{b}})Tufano, Watson, Bavota, Penta, White, and Poshyvanyk}]{tufanoEmpiricalStudyLearning2019}
Tufano M, Watson C, Bavota G, Penta MD, White M, Poshyvanyk D (2019{\natexlab{b}}) An {{Empirical Study}} on {{Learning Bug-Fixing Patches}} in the {{Wild}} via {{Neural Machine Translation}}. ACM Transactions on Software Engineering and Methodology 28(4):19:1--19:29, \doi{10.1145/3340544}

\bibitem[{Vacheret et~al(2024)Vacheret, P{\'e}rez, Ziadi, and Hillah}]{vacheretBoostingFaultLocalization2024}
Vacheret R, P{\'e}rez F, Ziadi T, Hillah L (2024) Boosting fault localization of statements by combining topic modeling and {{Ochiai}}. Information and Software Technology 173:107,499, \doi{10.1016/j.infsof.2024.107499}

\bibitem[{Wang et~al(2023)Wang, Wang, Joty, and Hoi}]{wangRAPGenRetrievalAugmentedPatch2023}
Wang W, Wang Y, Joty S, Hoi SC (2023) {{RAP-Gen}}: {{Retrieval-Augmented Patch Generation}} with {{CodeT5}} for {{Automatic Program Repair}}. In: Proceedings of the 31st {{ACM Joint European Software Engineering Conference}} and {{Symposium}} on the {{Foundations}} of {{Software Engineering}}, Association for Computing Machinery, New York, NY, USA, {{ESEC}}/{{FSE}} 2023, pp 146--158, \doi{10.1145/3611643.3616256}

\bibitem[{Wang et~al(2021)Wang, Wang, Joty, and Hoi}]{wangCodeT5IdentifierawareUnified2021}
Wang Y, Wang W, Joty S, Hoi SC (2021) {{CodeT5}}: {{Identifier-aware Unified Pre-trained Encoder-Decoder Models}} for {{Code Understanding}} and {{Generation}}. In: Proceedings of the 2021 {{Conference}} on {{Empirical Methods}} in {{Natural Language Processing}}, Association for Computational Linguistics, Online and Punta Cana, Dominican Republic, pp 8696--8708, \doi{10.18653/v1/2021.emnlp-main.685}

\bibitem[{Wei et~al(2023)Wei, Xia, and Zhang}]{weiCopilotingCopilotsFusing2023}
Wei Y, Xia CS, Zhang L (2023) Copiloting the {{Copilots}}: {{Fusing Large Language Models}} with {{Completion Engines}} for {{Automated Program Repair}}. In: Proceedings of the 31st {{ACM Joint European Software Engineering Conference}} and {{Symposium}} on the {{Foundations}} of {{Software Engineering}}, Association for Computing Machinery, New York, NY, USA, {{ESEC}}/{{FSE}} 2023, pp 172--184, \doi{10.1145/3611643.3616271}

\bibitem[{Wes(2010)}]{wesDataStructuresStatistical2010}
Wes M (2010) Data {{Structures}} for {{Statistical Computing}} in {{Python}}. In: Proceedings of the 9th {{Python}} in {{Science Conference}}, SciPy, vol 445, pp 56--61, \doi{10.25080/majora-92bf1922-00a}

\bibitem[{Widyasari et~al(2020)Widyasari, Sim, Lok, Qi, Phan, Tay, Tan, Wee, Tan, Yieh, Goh, Thung, Kang, Hoang, Lo, and Ouh}]{widyasariBugsInPyDatabaseExisting2020}
Widyasari R, Sim SQ, Lok C, Qi H, Phan J, Tay Q, Tan C, Wee F, Tan JE, Yieh Y, Goh B, Thung F, Kang HJ, Hoang T, Lo D, Ouh EL (2020) {{BugsInPy}}: A database of existing bugs in {{Python}} programs to enable controlled testing and debugging studies. In: Proceedings of the 28th {{ACM Joint Meeting}} on {{European Software Engineering Conference}} and {{Symposium}} on the {{Foundations}} of {{Software Engineering}}, Association for Computing Machinery, New York, NY, USA, {{ESEC}}/{{FSE}} 2020, pp 1556--1560, \doi{10.1145/3368089.3417943}

\bibitem[{Wolf et~al(2020)Wolf, Debut, Sanh, Chaumond, Delangue, Moi, Cistac, Rault, Louf, Funtowicz, Davison, Shleifer, {von Platen}, Ma, Jernite, Plu, Xu, Le~Scao, Gugger, Drame, Lhoest, and Rush}]{wolfTransformersStateoftheArtNatural2020}
Wolf T, Debut L, Sanh V, Chaumond J, Delangue C, Moi A, Cistac P, Rault T, Louf R, Funtowicz M, Davison J, Shleifer S, {von Platen} P, Ma C, Jernite Y, Plu J, Xu C, Le~Scao T, Gugger S, Drame M, Lhoest Q, Rush A (2020) Transformers: {{State-of-the-Art Natural Language Processing}}. In: Proceedings of the 2020 {{Conference}} on {{Empirical Methods}} in {{Natural Language Processing}}: {{System Demonstrations}}, Association for Computational Linguistics, Online, pp 38--45, \doi{10.18653/v1/2020.emnlp-demos.6}

\bibitem[{Xia and Zhang(2022)}]{xiaLessTrainingMore2022}
Xia CS, Zhang L (2022) Less training, more repairing please: Revisiting automated program repair via zero-shot learning. In: Proceedings of the 30th {{ACM Joint European Software Engineering Conference}} and {{Symposium}} on the {{Foundations}} of {{Software Engineering}}, Association for Computing Machinery, New York, NY, USA, {{ESEC}}/{{FSE}} 2022, pp 959--971, \doi{10.1145/3540250.3549101}

\bibitem[{Xia and Zhang(2024)}]{xiaAutomatedProgramRepair2024}
Xia CS, Zhang L (2024) Automated {{Program Repair}} via {{Conversation}}: {{Fixing}} 162 out of 337 {{Bugs}} for \$0.42 {{Each}} using {{ChatGPT}}. In: Proceedings of the 33rd {{ACM SIGSOFT International Symposium}} on {{Software Testing}} and {{Analysis}}, Association for Computing Machinery, New York, NY, USA, {{ISSTA}} 2024, pp 819--831, \doi{10.1145/3650212.3680323}

\bibitem[{Xia et~al(2023{\natexlab{a}})Xia, Ding, and Zhang}]{xiaPlasticSurgeryHypothesis2023}
Xia CS, Ding Y, Zhang L (2023{\natexlab{a}}) The {{Plastic Surgery Hypothesis}} in the {{Era}} of {{Large Language Models}}. In: 2023 38th {{IEEE}}/{{ACM International Conference}} on {{Automated Software Engineering}} ({{ASE}}), pp 522--534, \doi{10.1109/ASE56229.2023.00047}

\bibitem[{Xia et~al(2023{\natexlab{b}})Xia, Wei, and Zhang}]{xiaAutomatedProgramRepair2023}
Xia CS, Wei Y, Zhang L (2023{\natexlab{b}}) Automated {{Program Repair}} in the {{Era}} of {{Large Pre-trained Language Models}}. In: 2023 {{IEEE}}/{{ACM}} 45th {{International Conference}} on {{Software Engineering}} ({{ICSE}}), pp 1482--1494, \doi{10.1109/ICSE48619.2023.00129}

\bibitem[{Yang et~al(2021)Yang, Liu, Kim, Koyuncu, Kim, Tian, Lei, Mao, Klein, and Bissyand{\'e}}]{yangWhereWereRepair2021}
Yang D, Liu K, Kim D, Koyuncu A, Kim K, Tian H, Lei Y, Mao X, Klein J, Bissyand{\'e} TF (2021) Where were the repair ingredients for {{Defects4j}} bugs? Empirical Software Engineering 26(6):122, \doi{10.1007/s10664-021-10003-7}

\bibitem[{Yang et~al(2024)Yang, Jimenez, Wettig, Lieret, Yao, Narasimhan, and Press}]{yangSWEagentAgentComputerInterfaces2024}
Yang J, Jimenez CE, Wettig A, Lieret K, Yao S, Narasimhan K, Press O (2024) {{SWE-agent}}: {{Agent-Computer Interfaces Enable Automated Software Engineering}}. Advances in Neural Information Processing Systems 37:50,528--50,652, \url{https://proceedings.neurips.cc/paper_files/paper/2024/hash/5a7c947568c1b1328ccc5230172e1e7c-Abstract-Conference.html}

\bibitem[{Yasunaga and Liang(2020)}]{yasunagaGraphbasedSelfSupervisedProgram2020}
Yasunaga M, Liang P (2020) Graph-based, {{Self-Supervised Program Repair}} from {{Diagnostic Feedback}}. In: Proceedings of the 37th {{International Conference}} on {{Machine Learning}}, PMLR, pp 10,799--10,808, \url{https://proceedings.mlr.press/v119/yasunaga20a.html}

\bibitem[{Yasunaga and Liang(2021)}]{yasunagaBreakItFixItUnsupervisedLearning2021}
Yasunaga M, Liang P (2021) Break-{{It-Fix-It}}: {{Unsupervised Learning}} for {{Program Repair}}. In: Proceedings of the 38th {{International Conference}} on {{Machine Learning}}, PMLR, pp 11,941--11,952, \url{https://proceedings.mlr.press/v139/yasunaga21a.html}

\bibitem[{Ye and Monperrus(2024)}]{yeITERIterativeNeural2024}
Ye H, Monperrus M (2024) {{ITER}}: {{Iterative Neural Repair}} for {{Multi-Location Patches}}. In: Proceedings of the 46th {{IEEE}}/{{ACM International Conference}} on {{Software Engineering}}, Association for Computing Machinery, New York, NY, USA, {{ICSE}} '24, pp 1--13, \doi{10.1145/3597503.3623337}

\bibitem[{Ye et~al(2021{\natexlab{a}})Ye, Martinez, Durieux, and Monperrus}]{yeComprehensiveStudyAutomatic2021}
Ye H, Martinez M, Durieux T, Monperrus M (2021{\natexlab{a}}) A comprehensive study of automatic program repair on the {{QuixBugs}} benchmark. Journal of Systems and Software 171:110,825, \doi{10.1016/j.jss.2020.110825}

\bibitem[{Ye et~al(2021{\natexlab{b}})Ye, Martinez, and Monperrus}]{yeAutomatedPatchAssessment2021}
Ye H, Martinez M, Monperrus M (2021{\natexlab{b}}) Automated patch assessment for program repair at scale. Empirical Software Engineering 26(2):20, \doi{10.1007/s10664-020-09920-w}

\bibitem[{Ye et~al(2022)Ye, Martinez, and Monperrus}]{yeNeuralProgramRepair2022}
Ye H, Martinez M, Monperrus M (2022) Neural program repair with execution-based backpropagation. In: Proceedings of the 44th {{International Conference}} on {{Software Engineering}}, Association for Computing Machinery, New York, NY, USA, {{ICSE}} '22, pp 1506--1518, \doi{10.1145/3510003.3510222}

\bibitem[{Ye et~al(2023)Ye, Martinez, Luo, Zhang, and Monperrus}]{yeSelfAPRSelfsupervisedProgram2023}
Ye H, Martinez M, Luo X, Zhang T, Monperrus M (2023) {{SelfAPR}}: {{Self-supervised Program Repair}} with {{Test Execution Diagnostics}}. In: Proceedings of the 37th {{IEEE}}/{{ACM International Conference}} on {{Automated Software Engineering}}, Association for Computing Machinery, New York, NY, USA, {{ASE}} '22, pp 1--13, \doi{10.1145/3551349.3556926}

\bibitem[{Yi and Ismayilzada(2022)}]{yiSpeedingConstraintbasedProgram2022}
Yi J, Ismayilzada E (2022) Speeding up constraint-based program repair using a search-based technique. Information and Software Technology 146:106,865, \doi{10.1016/j.infsof.2022.106865}

\bibitem[{Yuan et~al(2022)Yuan, Zhang, He, Fang, Hung, Hao, and Yin}]{yuanCIRCLEContinualRepair2022}
Yuan W, Zhang Q, He T, Fang C, Hung NQV, Hao X, Yin H (2022) {{CIRCLE}}: Continual repair across programming languages. In: Proceedings of the 31st {{ACM SIGSOFT International Symposium}} on {{Software Testing}} and {{Analysis}}, Association for Computing Machinery, New York, NY, USA, {{ISSTA}} 2022, pp 678--690, \doi{10.1145/3533767.3534219}

\bibitem[{Zhang et~al(2023{\natexlab{a}})Zhang, Fang, Ma, Sun, and Chen}]{zhangSurveyLearningbasedAutomated2023}
Zhang Q, Fang C, Ma Y, Sun W, Chen Z (2023{\natexlab{a}}) A {{Survey}} of {{Learning-based Automated Program Repair}}. ACM Transactions on Software Engineering and Methodology 33(2):55:1--55:69, \doi{10.1145/3631974}

\bibitem[{Zhang et~al(2023{\natexlab{b}})Zhang, Fang, Zhang, Yu, Sun, and Chen}]{zhangGammaRevisitingTemplateBased2023}
Zhang Q, Fang C, Zhang T, Yu B, Sun W, Chen Z (2023{\natexlab{b}}) Gamma: {{Revisiting Template-Based Automated Program Repair Via Mask Prediction}}. In: 2023 38th {{IEEE}}/{{ACM International Conference}} on {{Automated Software Engineering}} ({{ASE}}), pp 535--547, \doi{10.1109/ASE56229.2023.00063}

\bibitem[{Zhang et~al(2024{\natexlab{a}})Zhang, Fang, Xie, Ma, Sun, Yang, and Chen}]{zhangSystematicLiteratureReview2024}
Zhang Q, Fang C, Xie Y, Ma Y, Sun W, Yang Y, Chen Z (2024{\natexlab{a}}) A {{Systematic Literature Review}} on {{Large Language Models}} for {{Automated Program Repair}}. \doi{10.48550/arXiv.2405.01466}, \eprint{2405.01466}

\bibitem[{Zhang et~al(2024{\natexlab{b}})Zhang, Zhang, Zhai, Fang, Yu, Sun, and Chen}]{zhangCriticalReviewLarge2024}
Zhang Q, Zhang T, Zhai J, Fang C, Yu B, Sun W, Chen Z (2024{\natexlab{b}}) A {{Critical Review}} of {{Large Language Model}} on {{Software Engineering}}: {{An Example}} from {{ChatGPT}} and {{Automated Program Repair}}. \doi{10.48550/arXiv.2310.08879}, \eprint{2310.08879}

\bibitem[{Zhang et~al(2023{\natexlab{c}})Zhang, Li, Jin, and Xing}]{zhangNeuralProgramRepair2023}
Zhang Y, Li G, Jin Z, Xing Y (2023{\natexlab{c}}) Neural {{Program Repair}} with {{Program Dependence Analysis}} and {{Effective Filter Mechanism}}. \doi{10.48550/arXiv.2305.09315}, \eprint{2305.09315}

\bibitem[{Zhang et~al(2024{\natexlab{c}})Zhang, Ruan, Fan, and Roychoudhury}]{zhangAutoCodeRoverAutonomousProgram2024}
Zhang Y, Ruan H, Fan Z, Roychoudhury A (2024{\natexlab{c}}) {{AutoCodeRover}}: {{Autonomous Program Improvement}}. In: Proceedings of the 33rd {{ACM SIGSOFT International Symposium}} on {{Software Testing}} and {{Analysis}}, Association for Computing Machinery, New York, NY, USA, {{ISSTA}} 2024, pp 1592--1604, \doi{10.1145/3650212.3680384}

\bibitem[{Zhong et~al(2023)Zhong, Ge, Ai, Li, Liu, Ge, and Luo}]{zhongStandUp4NPRStandardizingSetUp2023}
Zhong W, Ge H, Ai H, Li C, Liu K, Ge J, Luo B (2023) {{StandUp4NPR}}: {{Standardizing SetUp}} for {{Empirically Comparing Neural Program Repair Systems}}. In: Proceedings of the 37th {{IEEE}}/{{ACM International Conference}} on {{Automated Software Engineering}}, Association for Computing Machinery, New York, NY, USA, {{ASE}} '22, pp 1--13, \doi{10.1145/3551349.3556943}

\bibitem[{Zhong et~al(2024)Zhong, Li, Liu, Xu, Ge, Bissyande, Luo, and Ng}]{zhongPracticalProgramRepair2024}
Zhong W, Li C, Liu K, Xu T, Ge J, Bissyande TF, Luo B, Ng V (2024) Practical {{Program Repair}} via {{Preference-based Ensemble Strategy}}. In: Proceedings of the {{IEEE}}/{{ACM}} 46th {{International Conference}} on {{Software Engineering}}, Association for Computing Machinery, New York, NY, USA, {{ICSE}} '24, pp 1--13, \doi{10.1145/3597503.3623310}

\bibitem[{Zhu et~al(2021)Zhu, Sun, Xiao, Zhang, Yuan, Xiong, and Zhang}]{zhuSyntaxguidedEditDecoder2021}
Zhu Q, Sun Z, Xiao Ya, Zhang W, Yuan K, Xiong Y, Zhang L (2021) A syntax-guided edit decoder for neural program repair. In: Proceedings of the 29th {{ACM Joint Meeting}} on {{European Software Engineering Conference}} and {{Symposium}} on the {{Foundations}} of {{Software Engineering}}, Association for Computing Machinery, New York, NY, USA, {{ESEC}}/{{FSE}} 2021, pp 341--353, \doi{10.1145/3468264.3468544}

\bibitem[{Zhu et~al(2023)Zhu, Sun, Zhang, Xiong, and Zhang}]{zhuTareTypeAwareNeural2023}
Zhu Q, Sun Z, Zhang W, Xiong Y, Zhang L (2023) Tare: {{Type-Aware Neural Program Repair}}. In: 2023 {{IEEE}}/{{ACM}} 45th {{International Conference}} on {{Software Engineering}} ({{ICSE}}), pp 1443--1455, \doi{10.1109/ICSE48619.2023.00126}

\end{thebibliography}


\end{document}